\documentclass[twocolumn,aps,prb]{revtex4-1}
\usepackage{graphicx}
\usepackage{amsmath, bm}

\newcommand{\fn}{\sigma}
\newcommand{\sn}{{\sigma}}

\newcommand{\cH}{{\cal H}}

\newcommand{\cV}{{\cal V}}
\newcommand{\la}{\langle}
\newcommand{\ra}{\rangle}

\begin{document}

\newlength{\figurewidth}
\setlength{\figurewidth}{\columnwidth}

\title{Electronic structure and the glass transition in pnictide and
  chalcogenide semiconductor alloys. Part I: The formation of the
  $pp\sigma$-network.} \author{Andriy Zhugayevych$^1$}
\author{Vassiliy Lubchenko$^{1,2}$}\email{vas@uh.edu}
\affiliation{$^1$Department of Chemistry, University of Houston, TX
  77204-5003 \\ $^2$Department of Physics, University of Houston, TX
  77204-5005}
\begin{abstract}

  Semiconductor glasses exhibit many unique optical and electronic
  anomalies. We have put forth a semi-phenomenological scenario ({\it
    J. Chem. Phys.} {\bf 132}, 044508 (2010)) in which several of
  these anomalies arise from deep midgap electronic states residing on
  high-strain regions intrinsic to the activated transport above the
  glass transition. Here we demonstrate at the molecular level how
  this scenario is realized in an important class of semiconductor
  glasses, namely chalcogen and pnictogen containing alloys. Both the
  glass itself and the intrinsic electronic midgap states emerge as a
  result of the formation of a network composed of $\sigma$-bonded
  atomic $p$-orbitals that are only weakly hybridized. Despite a large
  number of weak bonds, these $pp\sigma$-networks are stable with
  respect to competing types of bonding, while exhibiting a high
  degree of structural degeneracy. The stability is rationalized with
  the help of a hereby proposed structural model, by which
  $pp\sigma$-networks are symmetry-broken and distorted versions of a
  high symmetry structure. The latter structure exhibits exact
  octahedral coordination and is fully covalently-bonded. The present
  approach provides a microscopic route to a fully consistent
  description of the electronic and structural excitations in vitreous
  semiconductors.

\end{abstract}

\maketitle

\section{Introduction}

The electronic and structural excitations in amorphous semiconductors,
and the interplay of these excitations, have evaded a self-consistent
first-principles description for decades. Amorphous
semiconductors\cite{Morigaki1999, Zallen1998, Feltz1993, Elliott1990}
are important both in applications, e.g., as phase change
materials,\cite{ISI:000250615400019, ISI:000261127100019} and from the
basic viewpoint. The electronic structure in a disordered lattice is
fundamentally different from the venerable Bloch picture of continuous
bands of allowed states separated by strictly forbidden gaps, as would
be applicable in periodic solids. Although a result of multiple
electron scattering, the presence of such forbidden gaps in periodic
solids is also consistent with their energetic stability in that such
gaps usually imply stabilized occupied orbitals. In contrast, in a
disordered lattice, strict gaps in the density of states are not
\textit{\`{a} priori} allowed. Yet the electronic orbitals can still
be subdivided in two relatively distinct classes:\cite{AndersonLoc,
  Mott1982, CFO, Mott1990} (a) extended states, in which electrons
move as well-defined wave-packets, and (b) localized states, whose
density decays nearly exponentially away from mobility bands. Despite
the many successes in applying these ideas semi-phenomenologically,
developing a first principles description, in which a realistically
stable aperiodic lattice and mobility gaps emerge
self-consistently,\cite{ISI:A1972L738100004, PWA_negU} has been
difficult.\cite{RevModPhys.50.203} Further compounding this
difficulty, several electronic and optical peculiarities of disordered
semiconductors indicate that there are effects of disorder beyond
those generically expected of a mechanically \emph{stable} disordered
lattice.

We have argued\cite{ZL_JCP} that, instead, it is not the aperiodicity
alone, but the intrinsic \emph{metastability} of semiconductor glasses
that leads to many unique properties of these disordered solids. A
glass is made by thermally quenching a supercooled liquid at a rate
exceeding the typical liquid relaxation rate.  A supercooled liquid
can be thought of as an aperiodic crystal characterized by myriad low
free energy configurations that are nearly degenerate. Molecular
motions in the liquid occur via local activated transitions between
these low free energy configurations,\cite{KTW, XW, LW_aging} which
are accompanied by the creation of high-strain interfacial regions
separating the configurations. The interfaces are {\em intrinsic} to
the activated dynamics; their concentration, for a given quenched
glass, depends (logarithmically slow) only on the time scale of the
glass transition, i.e., the quench speed.  According to the Random
First Order Transition (RFOT) theory of the glass
transition,\cite{KTW, XW, LW_aging, LW_ARPC} this concentration just
above the glass transition temperature on one hour scale, $T_g$, is
generically about $\xi^{-3} \simeq 10^{20}$~cm$^{-3}$, within an order
of magnitude or so, depending on the specific substance. The parameter
$\xi$ denotes the cooperativity length for activated reconfigurations.

Our results indicate\cite{ZL_JCP} that the high-strain regions that
form when the amorphous material is made may host deep midgap
electronic states of topological origin, which are centered on over-
and under-coordinated atoms.  These states share several
characteristics with the midgap electronic states in
trans-polyacetylene, which are centered on defects in the perfect
alternation pattern of the double and single bonds along the polymer
chains.\cite{RevModPhys.60.781} Using a semi-phenomenological,
coarse-grained Hamiltonian, we have established the spatial and charge
characteristics of the interface-based midgap states in non-polymeric
glasses.  We further concluded, based on the internal consistency of
the description, that the states should be present only in a limited
class of glasses that satisfy the following requirements: The bonding
should exhibit inhomogeneous saturation so that the transfer integrals
$t$ in the electronic effective tight-binding Hamiltonian $\cH =
\sum_i \epsilon_i c_i^\dagger c_i^{\phantom{\dagger}} + \sum_{\la ij
  \ra} t_{ij} c_i^\dagger c_j^{\phantom{\dagger}}$ should uniformly
exhibit spatial variation.  Nevertheless, the magnitude of the
variation should be modest: \cite{ZL_JCP}
\begin{equation} \label{tratio} |t'/t| \gtrsim 0.5,
\end{equation}
where $t$ and $t'$ denote the upper and lower limits of the variation
range. More detailed estimates\cite{ZLMicro2} indicate that the lower
limit on the $t'/t$ ratio is probably smaller, i.e., 0.3 or
so. Finally, the spatial variation $\delta \epsilon$ in
electronegativity should not be too large
\begin{equation} \label{enegConstr} |\delta \epsilon| < |t-t'|,
\end{equation}
thus implying the material is a semiconductor, since the transfer
integral $t$ is at most a few eV. Of non-polymeric materials, only
certain chalcogen- and pnictogen-containing glasses appear to satisfy
all of these requirements. On the other hand, these amorphous arsenic
chalcogenides and similar materials do indeed exhibit several
electronic and optical anomalies that could be accounted for by the
interface-based states, in a unified fashion.\cite{ZL_JCP} These
anomalies include light-induced electron spin resonance (ESR) and
midgap absorption,\cite{BiegelsenStreet, PhysRevB.38.11048} two types
of photoluminescence,\cite{TadaNinomiya} and field-induced ultrasonic
attenuation.\cite{ClaytorSladek} Thus general arguments, on the one
hand, and observation, on the other, seem to converge on the
uniqueness of chalcogenide and pnictide glasses with regard to their
potential ability to host topological midgap states. Despite this
remarkable convergence, the currently available evidence for the
unique interplay between electronic excitations and the metastability
in those glasses must be regarded as circumstantial.

\begin{figure}[t]
\centering
\includegraphics[width=0.6\figurewidth]{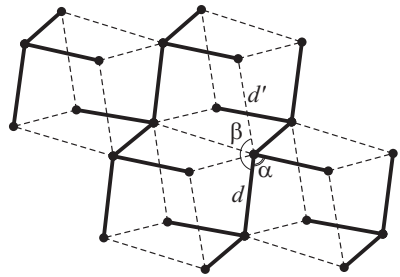}
\caption{Structure of rhombohedral arsenic as an example of a
  $pp\sigma$-bonded network.  The solid lines denote regular covalent
  bonds that connect the central atom with the nearest neighbors, bond
  length $d$. The dashed line denote the weaker, ``secondary'' bonds
  connecting the central atom with its next nearest neighbors, bond
  length $d'$. Angles $\beta < 180^\circ$ and $\alpha \ne 90^\circ$
  reflect the deviation from ideal octahedral coordination. These
  particular crystal fragment and view are based on Fig. 1 of Shang et
  al.\cite{PhysRevB.76.052301}}
\label{fig_coo}
\end{figure}

The purpose of the present effort is to test the conclusions of the
semi-phenomenological analysis from Ref.\cite{ZL_JCP} directly, based
on the local chemistry specific to amorphous chalcogenide and pnictide
alloys.  Our basic hypothesis for the origin of the topological midgap
states in the semiconductor glasses is that these glasses represent
aperiodic networks of $\sigma$-bonded $p$-orbitals that are only
weakly hybridized, as in Fig.~\ref{fig_coo}. These networks exhibit a
relatively small number of intrinsic over- and under-coordinated
vertices. The latter are, in fact, responsible for both the midgap
states and the transition state configurations intrinsic to molecular
transport in the quenched melts.  In testing this hypothesis, we face
the deeper question of the actual stability of aperiodic
$pp\sigma$-bonded networks. Experiment shows (see below), that the
enthalpy excess of a glass relative to the corresponding crystal is
\emph{less than the typical vibrational energy}, i.e., a small
fraction - one percent or so - of the total bonding energy! In other
words, despite their aperiodicity, glasses are nearly defect-free
structures, consistent with their bulk stability, both mechanical and
thermodynamic. This observation is in conflict with a common view of
glasses as a reconstructed - but otherwise arbitrary - array of
malcoordinated configurations and other local defects, such as
vacancies. This common view would imply excess enthalpies of the order
eV per several atoms, while avoiding to address the mechanism of
transport in the melt.

In the present and companion article,\cite{ZLMicro2} we test the
hereby proposed microscopic picture in two relatively separate
stages. The present article is devoted to the first stage, in which we
argue that $pp\sigma$-networks can represent the quenched liquid and
frozen glass forms of these substances in the first place. We will
make a case that (a) such networks are stable against other types of
bonding in a rather large class of amorphous compounds containing
elements from groups 15 and 16; and (b) despite their relative
stability, aperiodic $pp\sigma$ networks are multiply degenerate, as
are the actual liquids and glasses in question. The present work thus
contains, to our knowledge, the first chemical bonding theory of a
bulk glass.

The article is organized as follows.  In Section \ref{pbond}, we
discuss in detail several key features of $pp\sigma$ networks,
including their spatial non-uniformity and a hierarchy of bonding,
from strictly covalent to weaker ``secondary'' to weaker yet van der
Waals.  Despite the presence of such weak bonds, the $pp\sigma$
networks are stable. To trace the origin of this stability, we
formulate a structural model in Section \ref{model}, by which both
periodic and \emph{aperiodic} $pp\sigma$-networks are symmetry-broken
versions of a highly symmetric, strongly bonded structure. This view
is similar but distinct from the common view of many elemental solids
as Peierls distorted simple-cubic lattices.\cite{Burdett1995} The
symmetry breaking is driven by several competing interactions,
including in particular $sp$-mixing; these, nevertheless, are only
strong enough to perturb but not qualitatively modify the basic
$pp\sigma$ bonding.  We will verify that the resulting aperiodic
$pp\sigma$-bonded lattice satisfies the three requirements for the
existence of the topological states listed above. Importantly, this
lattice will be argued to exhibit multiply degenerate configurations
that differ by precise coordination of individual atoms, consistent
with the possibility of activated transport in the corresponding
quenched melts. The precise degree of degeneracy and the possibility
of activated transport are both intimately related to the questions of
the concentration of the corresponding transition-state configurations
in the melt and the accompanying electronic excitations. The latter
questions are analyzed in the companion article.\cite{ZLMicro2}

\section{$pp\sigma$-bonded semiconductors: the role of the secondary
  $pp\sigma$-interaction}
\label{pbond}

The goal of this Section is to provide a detailed description of
$pp\sigma$-networks in pnictides and chalcogenides, as arising from
sigma-bonding between $p$-orbitals that are only weakly
$sp$--hybridized. Such a description is necessitated by the lack of
systematic comparative studies of the electronic properties of
chalcogenide and pnictide alloys, even though their structure itself
has received much attention.\cite{Popescu00, Feltz1993} The
$pp\sigma$-bonding emerges subject to competition from other types of
local ordering. The presence of several competing types of local
ordering in chalcogenides is evidenced by their broad range of
structural and electronic properties, as could be seen by comparing
e.g.  As$_2$Se$_3$\cite{Li00} and GeSe$_2$.\cite{Cobb96} While the
former exhibits a distorted octahedral coordination and well separated
$s$ and $p$ bands, the latter displays tetrahedral ordering and
overlapping sets of $s$ and $p$ orbitals. At the same time, the two
substances exhibit opposite trends in terms of intrinsic and light
induced ESR response.\cite{BiegelsenStreet, PhysRevB.15.2278,
  Mollot1980, PhysRevB.38.11048} Such simultaneous trends are
particularly vivid in the Ge$_x$Se$_{1-x}$ series\cite{Salmon2007} for
$1/3 < x < 1/2$, which exhibit coordination ranging from tetrahedral
(smaller $x$) to distorted octahedral (larger $x$). In this series,
the octahedral ordering seems to correlate with the separation between
$s$ and $p$ bands\cite{Hachiya01, Makinistian07} and light induced
ESR, and anti-correlate with the presence of unpaired
spins\cite{Mollot1980} and the glassforming ability.\cite{Azoulay1975}
Vice versa, the tetrahedral bonding exhibits the opposite trend.

\begin{figure}[t]
  \centering \fbox{\includegraphics[width=0.6
    \figurewidth]{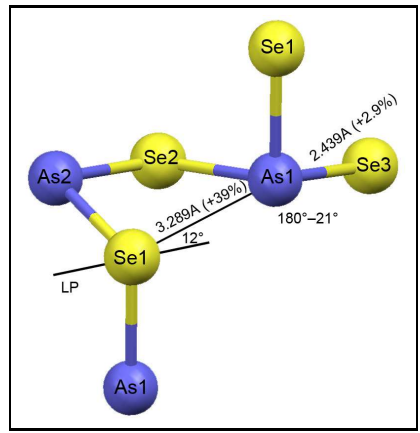}}
  \caption{A fragment of the As$_2$Se$_3$ crystal. The secondary bond
    between As1 and Se1 is shown with a thin solid line. The deviation
    from the strict octahedral coordination is reflected in different
    values of covalent  and secondary bond strengths: $t'_\sn/t_\fn=0.44 < 1$,
    a deviation from their strict alignment $\Delta \beta=180^\circ -
    \beta = 21^\circ$, and bond elongation relative to the sum of the
    covalent radii: $\delta d= + 0.07$ \AA, $\delta d'= + 0.9$
    \AA. The notations are the same as in Fig.~\ref{fig_coo}. }
  \label{fig_As2Se3frag}
\end{figure}

When $sp$-hybridization is weak, each atom exhibits a distorted
octahedral coordination, as exemplified in Fig.~\ref{fig_coo}: Two or
three nearest neighbors are situated at the distance of the regular
covalent bond in an almost right-angled geometry. Opposite to each of
these covalently bonded neighbors, there is an atom at a distance that
exceeds the sum of the corresponding covalent radii, but is closer
than the sum of the corresponding van der Waals
radii.\cite{Pearson1972} \emph{Crystalline} As, Se, As$_2$Se$_3$, GeSe
are typical examples of this type of coordination. It is appropriate
to think of crystals that exhibit distorted octahedral coordination
not as fully covalently bonded, but as networks consisting partially
of fully covalent bonds and weaker, closed-shell interactions. In the
physics literature, it is customary to call the stronger bonds ``front
bonds'' and to call the weaker bonds ``back bonds.''\cite{Harrison,
  Greaves79, Robertson83} Because the back bonds formally correspond
to closed-shell interactions, chemists call them ``secondary''
bonds,\cite{Alcock1972, Pyykko, LandrumHoffmann1998,
  PapoianHoffmann2000} or donor-acceptor interactions,\cite{Bent1968}
or, sometimes, hypervalent or 3-center bonds, where the distinction is
only quantitative, if any.\cite{LandrumHoffmann1998} Importantly, the
secondary bonds are stronger than van der Waals interaction and are
\emph{directional}, similarly to their strictly-covalent
counterparts.\cite{Alcock1972} We will see that, in fact, the
distinction between the secondary and covalent bonds in
$pp\sigma$-networks is not sharp.  A common example of the coexistence
of covalent and secondary bonding is crystalline As$_2$Se$_3$, which
consists of puckered layers of AsSe$_3$ pyramids, see also
Fig.~\ref{As2Se3Xtal}-\ref{linksSide} below, whereby the layers are
only loosely bonded. The pyramids are made of the stronger, covalent
bonds, while the secondary $pp\sigma$-interaction accounts for the
rest of the intralayer bonding, see Fig.~\ref{fig_As2Se3frag}.  The
interlayer secondary bonds are even weaker; nevertheless they have
been argued to be as strong as 0.3 eV in
As$_2$Se$_3$,\cite{PhysRevB.33.2968} i.e. significantly stronger than
a typical van der Waals-like bond.

Despite thousands of documented instances of secondary bonding and the
similarity in its properties across a broad spectrum of
compounds,\cite{LandrumHoffmann1998, Pyykko, Bent1968} both the
mechanism and quantitative description of this type of bonding still
appear to be a subject of debate.\cite{Pyykko} Without claiming full
generality, here we will presume the existence and universality of
secondary bonds based on those myriad documented cases. The energetic
and spatial characteristics of these bonds will be treated
empirically, using tight-binding (TB) formalism. Such an approach is
controllable insofar as the substances in question are insulators or
poor conductors since under these
circumstances,\cite{BurdettMetInsTrans} both localized, Wannier-like
and the delocalized Mulliken-Hund orbitals form a complete set of
electronic wave-functions. Consistent with the closed-shell character
of the secondary bonding, the TB expression for the corresponding
binding energy contains exclusively interaction between occupied and
unoccupied orbitals of the constituent molecules. Indeed, a
one-electron Hamiltonian that describes the interaction of two
molecules A and B can be written as a block matrix, where the ket
corresponding to the wave function is a vertical stack of the kets
pertaining to molecule A and B:
\begin{equation} \label{Hab}
  \cH=\begin{pmatrix} \cH^\text{A} & V^+ \\ V & \cH^\text{B} \end{pmatrix}, 
  | \psi \ra =\begin{pmatrix} | \psi^\text{A} \ra \\ | \psi^\text{B} 
    \ra \end{pmatrix}.
\end{equation}
Matrices $\cH^\text{A}$, $\cH^\text{B}$ are the Hamiltonians of the
isolated molecules A and B, while $V$ contains the corresponding
transfer integrals.  According to the standard perturbative
expression,\cite{MurrellKettleTedder} the binding energy of two
closed-shell molecules A and B, $E_\text{bind} \equiv
E_\text{tot}^\text{AB}-E_\text{tot}^\text{A}-E_\text{tot}^\text{B}$
reads:
\begin{equation} \label{Ebind} E_\text{bind} \approx 2
  \left\{\sum_{n}^{\text{A$_\text{occ}$}}\sum_{m}^{\text{B$_\text{unocc}$}}
    -\sum_{n}^{\text{A$_\text{unocc}$}}\sum_{m}^{\text{B$_\text{occ}$}}\right\}
  \frac{|V_{nm}|^2}{E_n^\text{A}-E_m^\text{B}},
\end{equation}
where $\cH^\text{A} |\psi_n^\text{A}\ra = E_n^\text{A}
|\psi_n^\text{A}\ra$ and $\cH^\text{B} |\psi_n^\text{B}\ra =
E_n^\text{B} |\psi_n^\text{B}\ra$. Labels ``occ'' (``unocc'') denote
summation over occupied (unoccupied) orbitals of the molecules A and
B.

\begin{table}[t]
  \centering
  \begin{tabular}{|c|ccccc|l|} \hline
    & $-t_{ss}$ & $t_{sp}$ & $t_\fn$ & $t'_\sn$ & $-t_\pi$ & ref. structure\\ \hline\hline
    Ge & 1.70 & 2.36 & 2.56 & \parbox{2em}{\centering\scriptsize negli\-gible} & 0.67 & diamond-Ge \\
    As & 1.17 & 1.60 & 3.10 & 1.66 & 0.79 & $\alpha$-As \\
    Se & 1.11 & 2.10 & 3.37 & 0.64 & 0.92 & trigonal-Se \\ \hline
  \end{tabular}
  \caption{Transfer integrals (in eV)  in particular crystalline forms
    of several elements often present in chalcogenide alloys.\cite{Robertson83}
    All integrals are for the nearest neighbors, except $t'_\sn$.
    In $t_{ss}$ and $t_{sp}$, the subscripts indicate the constituent orbitals.
    $t_\fn$ and $t'_\sn$ denote the transfer 
    integrals for the covalent and secondary  $pp\sigma$ bonds, and $t_\pi$ 
    for the $pp\pi$ interaction. A graphical summary of the transfer 
    integral definitions is given in Fig.~\ref{tdef}(a), see also p. 23 of 
    Ref.~\cite{Harrison}} 
  \label{tab.tint}
\end{table}

Now, the precise geometry of the $pp\sigma$-network is subject to
several competing interactions, combined with the precise
stoichiometry and other many body effects: the $pp\sigma$-interaction,
$sp$-mixing, and $pp\pi$-interaction, such as often found in
conjugated polymers. (See Harrison\cite{Harrison} for an introduction
to tight-binding methods.)  All these interactions have comparable
strength as can be inferred from the values of the corresponding
electron transfer integrals. Table~\ref{tab.tint} compiles the values
of these transfer integrals for important representatives from groups
fourteen, fifteen, and sixteen. Elements of these groups are of
particular interest in the context of amorphous semiconductors,
because of comparable electronegativity and suitable valency, of
course.  The close magnitude of the listed competing interactions
implies that the local order, which in turn is strongly affected by
the stoichiometry, plays the crucial role in determining which
interaction will ultimately dominate.

One may list several complementary ways to establish the presence and
significance of $pp\sigma$-bonding. A $pp\sigma$-network reveals
itself structurally in a weak deviation from the octahedral
coordination.  In addition to a nearly right-angled geometry, the
disparity between the lengths of the secondary and covalent bonds
should be modest.  In the latter case, the ratio of the corresponding
transfer integrals, $t'_\sn/t_\fn$, is not too small, implying a
relatively uniform, stable network. Furthermore, in view of a nearly
universal relation $-t_\pi/t_\fn \simeq 1/4$, Ref.~\cite{Harrison04},
a large enough value of $t'_\sn/t_\fn$ automatically guarantees that
the effect of $pp\pi$ interactions on the geometry is small. On the
other hand, when $sp$ mixing is weak and little $ss$ bonding is
present, the top of the valence band consists primarily of
$p$-orbitals, while the $s$ and $p$ subbands are relatively well
separated. Indeed, consider for the sake of argument two identical
centers, each having one $s$ and $p$ orbital and three electrons. The
$p$-orbitals are aligned. Within the second order in the $sp$-mixing,
the one-electron energies of the four resulting molecular orbitals are
given by $\epsilon_s\mp
t_{ss}-t_{sp}^2[\epsilon_p-\epsilon_s\pm(t_\sigma+t_{ss})]^{-1}$ for
the $ss\sigma$ bond and $\epsilon_p\pm t_\sigma+ t_{sp}^2 [\epsilon_p
- \epsilon_s\pm(t_\sigma+t_{ss})]^{-1}$ for the $pp\sigma$ bond. If
the $s$ and $p$ orbitals are sufficiently separated in energy, it
follows automatically that (a) the centers are $pp\sigma$-bonded and
(b) the effect of the $sp$-mixing on the $pp\sigma$ transfer integral
of the bond is small: $t_{sp}/ \sqrt{t_\sigma (\epsilon_p-\epsilon_s)}
< 1$.  A combination of the photoemission spectra with tight-binding
calculations, using the known crystal structures, strongly suggests
that exactly this type of bonding occurs in the crystals of several
archetypal chalcogenides: As$_2$S$_3$, As$_2$Se$_3$, and
As$_2$Te$_3$.\cite{PhysRevB.12.1567}

\begin{table}
\begin{tabular}{|lr|cc|ccc|c|}
\hline
\multicolumn{2}{|c|}{crystal and Ref.} & $\bar{\alpha},^\circ$ & $\epsilon$,eV & $\Delta d$,\AA & $\Delta d'$,\AA & $\Delta\beta,^\circ$ & $t'_\sn/t_\fn$ \\
\hline
\multicolumn{8}{l}{\quad strong $pp\sigma$-secondary bonding/hypervalency} \\
\hline
As$_2$Te$_3$   & \cite{Stergiou85a} &  90.8 & 0.1 &  0.1-0.4  & 0.4-1.2 & 3-21 & .4-.9 \\
r-As           & \cite{Schiferl69}  &  96.7 &  0  &    0.12   & 0.7 & 16 & .65 \\
\hline
\multicolumn{8}{l}{\quad moderate $pp\sigma$-secondary bonding} \\
\hline
o-As           & \cite{Smith74}     &  97.0 &  0  &    0.09   & 0.9 & 14 & .5 \\
GeSe           & \cite{Wiedemeier78}&  97.0 & 0.7 &    0.18   & 0.9 & 17 & .5 \\
t-Se at 5\,GPa & \cite{Keller77}    & 104.7 &  0  &    0.05   & 0.8 & 15 & .45 \\
As$_2$Se$_3$   & \cite{Stergiou85}  &  97.5 & 0.3 & 0.02-0.08 & 0.9-1.4 & 10-20 & .2-.4 \\
As$_4$Se$_4$   & \cite{Renninger73} &  98.4 & 0.3 &-0.04+0.11 & 1.1-1.5 &  2-15 & .2-.4 \\
t-Se at 0\,GPa & \cite{Keller77}    & 103.1 &  0  &    0.03   & 1.1 & 19 & .3 \\
r-Se           & \cite{Miyamoto80}  & 101.1 &  0  &    0.02   & 1.1 &  9 & .3 \\
Br at 5\,K     & \cite{Powell84}    &  --   &  0  &    0.02   & 1.0 & 10 & .3 \\
Br at 250\,K   & \cite{Powell84}    &  --   &  0  &    0.01   & 1.1 & 10 & .3 \\
$\alpha$-m-Se  & \cite{Cherin72}    & 105.6 &  0  &    0.00   & $\gtrsim 1.1$ & -- & $\lesssim .2$ \\
\hline
\hline
As$_2$O$_3$ (As) & \cite{Pertlik78} &  96.5 & 1.1 &  0.0-0.1  & 1.2 & 23 & .3 \\
AsBr$_3$ (As)  & \cite{Braekken35}  &  99.0 & 1.1 & -0.1+0.1  & 1.6 & 11 & .2 \\
\hline
\multicolumn{8}{l}{\quad $sp^3$-bonding} \\
\hline
GeSe$_2$ (Ge)  & \cite{Dittmar76}   & 109.4 & 0.7 &-0.06-0.03 & $\gtrsim 1.3$ &  8 & $<.3$ \\
h-Ge           & \cite{Zhang00}     & 109.4 &  0  &   -0.03   & 1.6 &  0 & .2 \\
diamond-Ge     & \cite{Wyckoff63}   & 109.5 &  0  &     0     & 2.2 & 30 & .1 \\
\hline
\end{tabular}
\caption{
  Geometry (notations from Fig.~\ref{fig_coo}) and tight-binding
  parameters  of  representative compounds in crystalline form. Lower entries
  correspond to progressively weaker $pp\sigma$-bonding.
  $\bar{\alpha}$ is the average bond angle (hybridization),
  $\epsilon$ is the half-difference in absolute electronegativities, 
  $\Delta d$ and $\Delta
  d'$ are the deviations of covalent and secondary bond lengths from the sum of
  the covalent radii, $\Delta\beta=180-\beta$ is the deviation from a
  linear geometry, $t'_\sn/t_\fn$ is the ratio of $pp\sigma$ integrals
  for covalent and secondary bonds. Crystallographic
  abbreviations: h -- hexagonal, m -- monoclinic, o -- orthorhombic, r --
  rhombohedral, t -- trigonal. The computational details  are
  given in Appendix \ref{app.tint}.
}
\label{tab.coo}
\end{table}

In Table~\ref{tab.coo}, we compile data on the deviation from the
ideal octahedral coordination and the corresponding $pp\sigma$
transfer integrals, in several distinct compositions and
stoichiometries characteristic of common chalcogenide and pnictide
alloys.  One observes that a high value of $t'_\sn/t_\fn$ is indeed
characteristic of $pp\sigma$-bonded materials, whereby the angular
deviation from the ideal coordination does not exceed $10^\circ$.
Conversely, a large value of the $t'_\sn/t_\fn$ ratio, alone, is often
a good predictor of $pp\sigma$-networking.  Note also a subtle, but
nevertheless significant trend that in such a network, the stronger
bonds are somewhat \emph{longer} than the sum of the covalent
radii. Furthermore, this deviation is the more significant, the
shorter - and hence stronger - are the secondary bonds. This
anti-correlation is a telltale sign of
``trans-influence''\cite{LandrumHoffmann1998} in which the weaker
bonded atom donates electrons into anti-bonding orbitals of the
stronger bond, a common feature with secondary and donor-acceptor
interactions. For the trans-influence to take place, it is essential
that the counterpart covalent and secondary bonds be in a near linear
geometry so that the anti-bonding orbital on the stronger bond overlap
significantly with the bonding orbital on the secondary
bond.\cite{Alcock1972} Note that tight-binding descriptions are
consistent with the trans-influence. Indeed, we show in Appendix B
that the bond order\cite{Armstrong73} for the AB dimer from
Eq.~(\ref{Ebind}) is approximately given by the expression:
\begin{equation} \label{BondOrder} b_\text{AB}
  \approx 
  4 \left\{ \sum_{n}^{\text{A$_\text{occ}$}} \sum_{m}^{\text{B$_\text{unocc}$}}
    + \sum_{n}^{\text{A$_\text{unocc}$}}\sum_{m}^{\text{B$_\text{occ}$}}\right\} 
  \frac{|V_{nm}|^2}{\left(E_n^\text{A}-E_m^\text{B}\right)^2}.
\end{equation}
The first (second) double sum is twice the occupation of the
antibonding orbitals of molecule B (A), as donated by the bonding
orbitals of A (B), consistent with the donor-acceptor nature of the
secondary bond.  Alternatively, according to Eq.~(\ref{Ebind}), the
bond order strongly correlates with the binding energy. Since the sum
of the bond orders on a given atom equals to the atom's valency,
stronger secondary bonds imply weaker counterpart covalent bonds.

Above said, we should point out that a large value of the
$t'_\sn/t_\fn$ ratio and the perfect octahedral coordination,
separately or together, do not guarantee that the $pp\sigma$-bonding
is the main contributor to the lattice stabilization. An obvious
counterexample is provided by ionic compounds with the rocksalt
structure, which exhibit the perfect octahedral coordination.
Incidentally, using stoichiometry to impose a (distorted) rocksalt
structure is not guaranteed to produce a $pp\sigma$-network
either. For instance, whereas GeSe does indeed exhibit distorted
octahedral coordination, AsSe forms a molecular crystal composed of
As$_4$Se$_4$ units, whose symmetry is incompatible with uniform
octahedral coordination, despite relatively strong secondary bonding, see
Table~\ref{tab.coo} and the Supplemental Material.\cite{SupplMat}
Finally, other competing types of local order are present in solids
formed by the elements from groups 14-16. For instance, in GeSe$_2$
the coordination of Ge atoms is tetrahedral, see Table~\ref{tab.coo}.
Elemental phosphorus and sulfur at ambient conditions exist as
molecular crystals made of tetraphosphorus P$_4$ and octasulphur
S$_8$, respectively, that show no signs of octahedral coordination.

\begin{figure}[t]
\centering
\includegraphics[width= 0.6 \figurewidth]{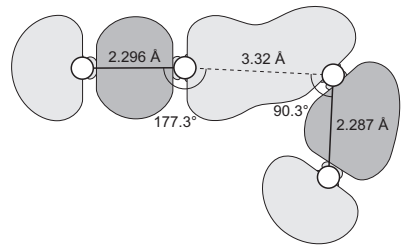}
\caption{A dimer of Br$_2$ molecules. The geometry is optimized at the
  MP2 level, using the program Firefly\cite{Firefly} with aug-cc-pVTZ
  basis set and RHF wave-function. The gray shapes show the lowest
  energy molecular orbital consisting of the valence $p$ orbitals of
  the Br$_2$ molecules (61st MO, Table~\ref{tab.dimerMO} in Appendix
  \ref{app.dimer}), see also the MO diagram in Fig.~\ref{BrMO}. The
  two hues reflect the sign of the wave function. The computational
  details are provided in Appendix \ref{app.dimer}.}
\label{fig_Brdimer}
\end{figure}

Apart from the stoichiometry and peculiar types of local ordering, the
actual degree of stabilization of the $pp\sigma$-network crucially
depends on the strength of the secondary bonds since they account for
at least a half of the total bonds.  We are not aware of systematic
\emph{ab initio} studies of the dependence of the strength of these
bonds on the bond length and the deviation from the precise octahedral
coordination, in a crystal.  Such studies are understandably
difficult, as the actual chalcogenide crystal structures exhibiting
this type of bonding are very complicated. For instance, the unit cell
of As$_2$Se$_3$ has 20 atoms.\cite{PhysRevB.33.2968} Despite these
complications, it is possible to obtain an accurate estimate of the
strength of $pp\sigma$ secondary bonding in semiconductors by
analyzing the simplest possible micro- and macro-molecular systems
that exhibit this type of bonding, i.e.  dimers of diatomic halogen
molecules and halogen crystals respectively. Specifically, bromine is
an appropriate example, because the elements of interest are located
in periods 3 through 5.  The ground state geometry of the bromine
dimer is shown in Fig.~\ref{fig_Brdimer}. Here the $pp\sigma$
molecular orbital of the l.h.s molecule is mixed with the in-plane
$pp\pi$ molecular orbitals of the r.h.s. molecule, as can be seen
directly in the electronic density distribution in
Fig.~\ref{fig_Brdimer}. Consistent with this mixing, the strength of
the $pp\sigma$ secondary bond between the bromine molecules exceeds
0.3 eV, i.e., significantly more than expected of a typical van der
Waals bond. According to the estimates on a variety of pseudo-dimer
structures by Anderson et al.\cite{ISI:A1994PH53100050}, the strength
of the secondary bonding, relative to the covalent bonding, decreases
toward the r.h.s. in each period. The above figure for the binding
energy of the bromine dimer thus gives us a secure lower bound on the
strength of a secondary bond, consistent with the data in
Table~\ref{tab.coo}.  Furthermore, bromine \emph{crystals} are
comprised of layers in which Br$_2$ units are arranged in nearly the
same geometry as in the ground state of an isolated pair of Br$_2$
molecules, see Supplementary Material.\cite{SupplMat}

\section{$pp\sigma$-networks as symmetry broken states}
\label{model}

Here we propose a specific structural model of $pp\sigma$-network
formation, both periodic and aperiodic, and argue that aperiodic
$pp\sigma$-networks are consistent with (a) the structural degeneracy
of the corresponding solid and (b) the restrictions on the magnitude
of the spatial variation of the electronic transfer integral from
Eq.~(\ref{tratio}) derived in Ref.\cite{ZL_JCP}

\begin{figure}[t]
\centering
\includegraphics[width=.6 \figurewidth]{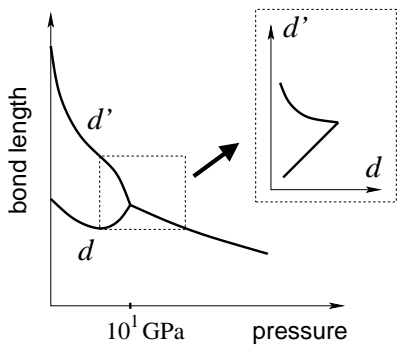}
\caption{Pressure dependence of the lengths of the covalent ($d$) and
  secondary ($d'$) bond in rhombohedral arsenic, after Silas et
  al.\cite{Silas08} At a critical pressure, of about $10^1$ GPa, the
  lattice becomes simple cubic, while the two types of bonds become
  equivalent. In the inset, we replot the framed region with $d$ as a
  function of $d'$ and the pressure as a dummy parameter. The
  resulting dependence illustrates the trans-influence of the covalent
  and secondary bonds, c.f. Figs. 1 and 2 of Landrum and
  Hoffmann.\cite{LandrumHoffmann1998}}
\label{symmetrybreaking}
\end{figure}

In liquids and glasses, the first coordination shell is determined by
the stronger bonds and is very similar to the first coordination shell
in the corresponding crystals.\cite{Bellissent87, Bellissent80,
  Hosokawa99} Much less is known about the precise configurations of
the weaker-bonded, next-nearest neighbors. A useful cue is provided by
the observation that the crystalline photoemission spectra of several
archetypal $pp\sigma$-bonded chalcogenides - As$_2$S$_3$,
As$_2$Se$_3$, and As$_2$Te$_3$\cite{PhysRevLett.26.1564,
  PhysRevB.12.1567} are very similar to their amorphous counterparts.
In this and the following Sections, we argue that, indeed, the local
interactions specific to $pp\sigma$-bonded glasses are of the same
origin as in the corresponding crystals: In both cases, the local
structures result from a symmetry-lowering transition from the perfect
octahedral coordination and thus are comparably stable. The main
corollary of this inference is that a supercooled liquid or quenched
glass can be sufficiently stabilized by the $pp\sigma$-network
alone. The key distinction between the crystal and glass is that owing
to its aperiodicity, the glass is necessarily structurally degenerate.

Let us begin with crystals. It is long
appreciated\cite{PhysRev.137.A871, Pearson1972, Takumi07,Fujihisa95}
that the structures of many polymorphs of elemental pnictogens,
chalcogens, and halogens can be regarded as distorted simple-cubic
(sc), with the exception of several of the lightest elements, such as
nitrogen or oxygen, which form molecular crystals.  Upon increasing
pressure to several tens of GPa, the structures approach the ideal
octahedral coordination, while phosphorus and arsenic actually exhibit
a continuous transition to the simple cubic structure
(Refs.\cite{PhysRevB.77.024109} and references therein). Specifically
in arsenic, which is rhombohedral (A7) at ambient conditions, as the
bond angles approach the right-angled geometry, the ratio of bond
lengths of the nearest to the next nearest neighbor
grows.\cite{Silas08} These two bond types correspond to the covalent
and secondary bonds respectively.  In the vicinity of the transition,
the covalent bond \emph{increases} in length, while the secondary bond
continues to shorten,\cite{Silas08} see Fig.~\ref{symmetrybreaking}.
Landrum and Hoffmann provide correlation data on thousands of
pnictogen and chalcogen compounds that clearly demonstrate similar
trans-influence between the covalent and secondary
bonds,\cite{LandrumHoffmann1998} see inset of
Fig.~\ref{symmetrybreaking}. The molecular fragments in question are
of the type X-Q-X, where Q=Sb, Te and X = F, Cl, Br, I. Note the
correlation in the inset of Fig.~\ref{symmetrybreaking} pertains to
valencies 3 (2) for Sb (Te). The combined view of
Fig.~\ref{symmetrybreaking} implies that the distinction between the
covalent and secondary bonds is not sharp, but is subject to precise
local coordination and/or bond tension.  In the strict sc limit, both
the secondary and covalent bonds become equivalent and should be
regarded as fully covalent, albeit
hypervalent.\cite{PapoianHoffmann2000} Similarly in halide crystals,
each $pp\sigma$-bonded layer transforms into a square lattice at a
sufficiently high pressure (80\,GPa for bromine\cite{Fujihisa95}).
Isovalent binary compounds A$^\text{IV}$B$^\text{VI}$ transform into
the simple cubic structure not only upon increasing pressure, but also
temperature.\cite{Littlewood80}

\begin{figure}[t]
    \includegraphics[width= .9 \figurewidth]{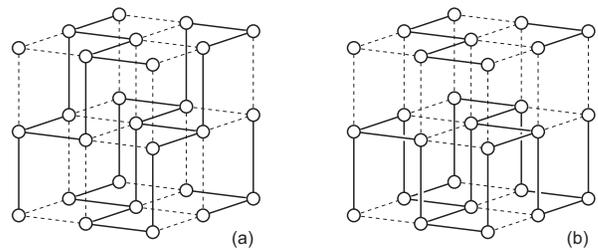}
    \caption{ \label{Burdett} Parent structures of the crystals of (a)
      elemental arsenic and (b) black phosphorus, after Burdett and
      McLarnan.\cite{burdett5764}}
\end{figure}

\begin{figure}[t]
  \centering
  \includegraphics[width=0.9\figurewidth]{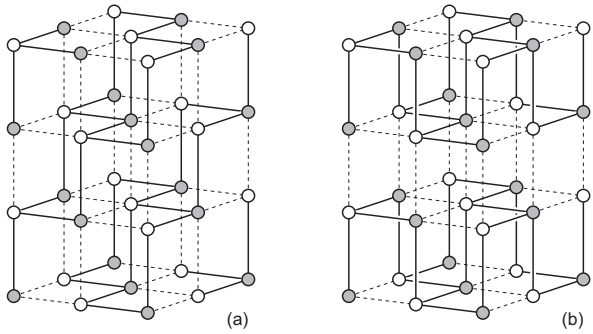}
  \caption{ \label{GeTeXtal} Parent structure for (a) the GeTe crystal
    and (b) GeSe and GeS crystals, c.f. Fig.~\ref{Burdett}. Note these
    structures are by one horizontal layer taller than the repeat
    unit.}
\end{figure}

To rationalize these observations, we hereby propose the following
\emph{structural model}, which draws heavily on Burdett and coworkers'
view of the structure of rhombohedral arsenic, black phosphorus and
other compounds.\cite{burdett5764, burdett5774} Place the atoms at the
vertices of the cubic lattice, so that each atom is exactly
octahedrally coordinated and the $p$ orbitals are aligned with the
principal axes. All atoms are linked, the links corresponding to
bonds. Remove links so that each atom obeys the octet rule, while
making sure the remaining bonds on each vertex are at 90 degrees, not
180. This procedure could be interpreted as adding electrons to a
rocksalt-like compound while breaking bonds, whereby each filled
antibonding orbital transforms into a lone pair of electrons pointing
away from the remaining bonds.\cite{burdett5764} As a result, each
pnictogen and chalcogen, for instance, will be three- and
two-coordinated respectively, whereby all links pointing from an atom
are at right angles.  We call the resulting lattice the ``parent
structure.''  Third, estimate the energy of the resulting parent
structure, using a tight-binding Hamiltonian, while assuming that the
transfer integrals are significant only for the linked atoms. Now,
those bond-breaking patterns that have a particular low energy are
special. One should expect that to these special structures, there
correspond crystals of actual substances that exhibit a distorted
octahedral coordination, in which the covalent bonds will precisely
correspond to the links, while the missing links correspond to
secondary bonds or weaker, van der Waals interactions. For instance,
Burdett and McLarnan have shown there are 36 inequivalent ways to
arrange three-coordinated atoms on the cubic lattice with a repeat
unit of size $2 \times 2 \times 2$. Two of the structures correspond
to the lattices of black phosphorus and rhombohedral
arsenic.\cite{burdett5764} In actual materials, both of these lattices
consist of double layers that are buckled and mutually shifted,
compared with the parent simple cubic structure.  Other specific
examples can be found in Refs.\cite{burdett5764, burdett5774,
  burdett1434} It is understood that although the simple cubic lattice
is a convenient parent structure for many compounds, it is by no means
unique in this regard. For instance, Albright et al.\cite{ABW} mention
two additional \emph{formal} ways to obtain the arsenic structure,
i.e. by adding two electrons per atom to wurzite ZnS or by puckering
graphite sheets.  Yet what distinguishes the sc-like parent structure
is that, like the actual material, it is $pp\sigma$-bonded, whereas
the orbitals in the wurzite and graphite structures are $sp^3$ and
$sp^2$ hybridized respectively.  We point out that IV-VI compounds
that are isoelectronic with arsenic can be obtained from the parent
structures of arsenic or phosphorus, see Fig.~\ref{GeTeXtal}. Finally
note that the above rules for bond placement, i.e. three-coordinated
pnictogens and two-coordinated chalcogens with right angles between
bonds, can be formally regarded as a subcase of the 64-vertex
model,\cite{0305-4470-22-2-003} which is the 3D generalization of the
venerable 6-vertex model of ice and 8-vertex model of
anti-ferroelectrics.\cite{Baxter} In the present case, 8
configurations on pnictogen vertices and 12 configurations on
chalcogen vertices have finite energies, while the rest are infinitely
costly. This analogy implies the proposed model is generalizable to
more complicated coordinations by assigning finite energies to the
latter.

\begin{figure}[t]
  \centering
  \includegraphics[width=0.9\figurewidth]{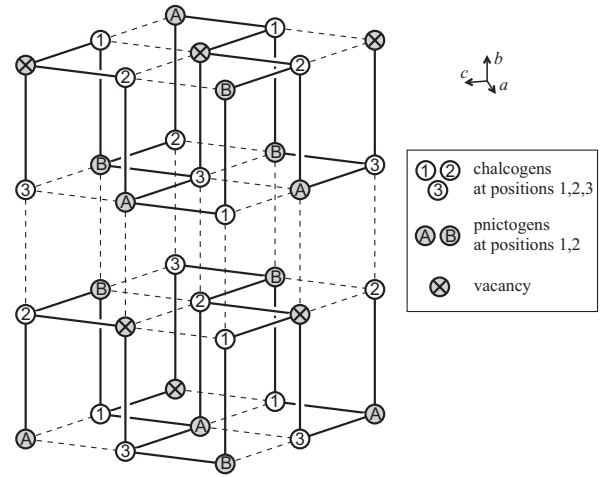}
  \caption{ \label{As2Se3Xtal} A parent structure for a Pn$_2$Ch$_3$
    crystal, such as crystalline As$_2$Se$_3$ and As$_2$S$_3$,
    c.f. Fig.~\ref{GeTeXtal}(b).}
\end{figure}

The present structural model can be formulated for stoichiometries
that can \emph{not} be arranged on the simple cubic lattice, except if
one allows for vacancies. A specific example of particular relevance
for this work is archetypal pnictogen-chalcogenides of stoichiometry
Pn$_2$Ch$_3$, such as As$_2$Se$_3$, that can form both a glass and a
crystal. (Pn = ``pnictogen,'' Ch = ``chalcogen.'')  In
Fig.~\ref{As2Se3Xtal} we show that coordination-wise and
symmetry-wise, the structure of this compound consists of double
layers similar to those in the black-phosphorus parent structure from
Fig.~\ref{Burdett}(b).  Note that by placing the vacancies in a
particular fashion, we achieve two things simultaneously: On the one
hand, the As$_2$Se$_3$ stoichiometry is obeyed, and on the other hand,
the octet rule on both pnictogens and chalcogens is satisfied.
In drawing individual double layers, we have used the structural model
of Vanderbilt and Joannopoulos,\cite{PhysRevB.23.2596} see also
Ref.\cite{ShimoiFukutome1} Note that based on the actual density of
As$_2$Se$_3$, the bond length in the parent structure in
Fig.~\ref{As2Se3Xtal} would have to be $2.8$ \AA, a reasonable number
consistent with the presence of the $pp\sigma$ network in the deformed
structure. The presence of vacancies in parent structures should not
be too surprising: The archetypal phase-change material
Ge$_2$Sb$_2$Te$_5$ is known to exhibit a (metastable) distorted cubic
structure with vacancies.\cite{Kolobov2004, Steimer2008}

\begin{figure}[t]
  \centering
  \includegraphics[width= .95 \figurewidth]{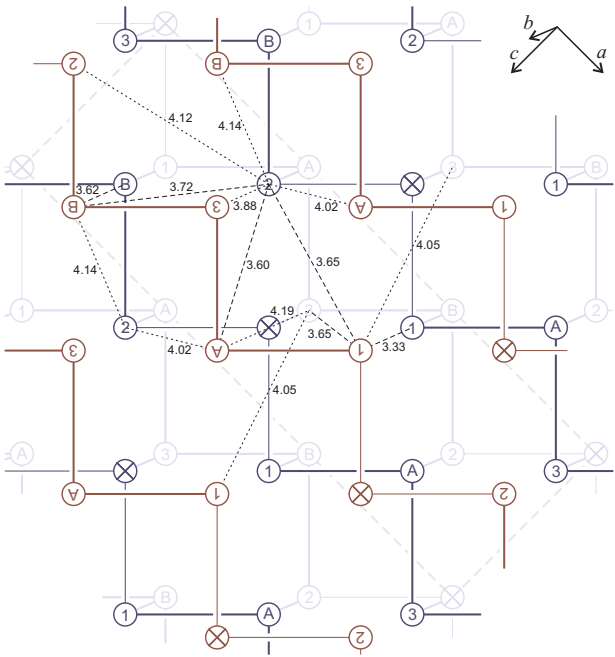}
  \caption{ \label{linksTop} The top view of a portion of the
    Pn$_2$Ch$_3$ parent structure from Fig.~\ref{As2Se3Xtal}. The
    labels of the top layer are upside-down, to help distinguish it
    from the bottom double-layer and indicate that adjacent double
    layers are related by inversion. The thin dashed lines and
    adjacent numbers indicate the corresponding secondary bonds and their
    length in the deformed structure, see Fig.~\ref{linksSide}. The
    tilted rectangle drawn with dashed lines indicates the unit
    cell. }
\end{figure}

\begin{figure}[t]
  \centering
  \includegraphics[width= \figurewidth]{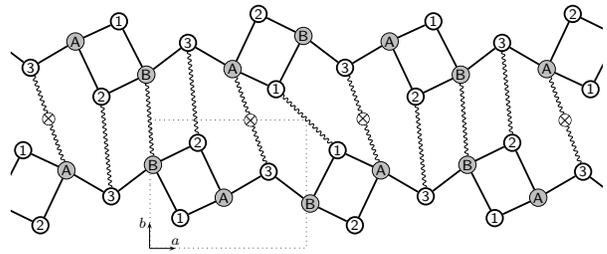}
  \caption{ \label{linksSide} A side view of the actual As$_2$Se$_3$
    structure. The wavy lines indicate inter-layer nearest neighbors
    in the parent structure, except the A-3 link, which is through a
    vacancy. Note that the bonded intra-layer atoms are automatically
    nearest neighbors in the parent structure. The lengths of the
    links are only partially indicative of the actual bond length
    because the bonds are not parallel to the projection plane.}
\end{figure}

The parent structure in Fig.~\ref{As2Se3Xtal} is clearly not unique in
that we could have placed the double layers in several distinct
positions relative to each other. The specific, ``homopolar''
arrangement in Fig.~\ref{As2Se3Xtal} was chosen because the sets of
close neighbors in this arrangement and in the actual deformed
structure seem to exhibit the greatest overlap, see
Figs.~\ref{linksTop} and \ref{linksSide}.  Nevertheless, several other
mutual positions are possible, which exhibit comparably overlapping
sets of close neighbors with the deformed structure, and, in addition,
minimize the distance between the vacancies in the parent structure
better than the specific realization in Figs.~\ref{As2Se3Xtal} and
\ref{linksTop}. For instance, consider shifting the top layer in
Fig.~\ref{linksTop} ``north'' by one lattice spacing.  Incidentally,
one notices that the vacancies in the parent structure become
``smeared'' in the interlayer space of the distorted structure, see
Supplementary Material.\cite{SupplMat} We point out that it would be
impossible to ``merge'' vacancies in the Pn$_2$Ch$_3$ stoichiometry
between \emph{each} two double-layers, if we attempted to use an
arsenic-like structure as the parent structure from
Fig.~\ref{Burdett}(a) instead of the black-phosphorus structure from
Fig.~\ref{Burdett}(b). This notion is consistent with the stability of
the latter structure in the actual material. At any rate, during the
distortion, the distance between \emph{linked} atoms decreases,
resulting in strong, covalent bonds. Conversely, the secondary bonds
will result partially from the cleaved links (usually intralayer) or
new contacts that formed in the deformed structure.
Figs.~\ref{As2Se3Xtal}-\ref{linksSide} indicate that because the
parent structure is not unique, there is no one-to-one correspondence
between missing links in the parent structure and the secondary bonds
in the distorted structure.

It is obvious that in the process of cleaving the bonds in the sc
structure, symmetry was lowered resulting in a twice bigger unit cell
along the pertinent directions. For example, in drawing the structure
in Fig.~\ref{As2Se3Xtal}, we could have split the original sc lattice
in double-layers in six equivalent ways, i.e., two along each
coordinate axis. Likewise, there are two equivalent ways to buckle
each double layer, in each of the (1,1) and (1,-1) directions. Further
symmetry lowering occurs when we place pnictogens, chalcogens and
vacancies at the lattice vertices. While the presence of symmetry
breaking itself is hereby obvious, its mechanism appears to be
subtler.  Several workers have argued the symmetry-breaking transition
that results in the structures of arsenic and phosphorus is
Peierls-like,\cite{Littlewood80, BurdettLeePeierls, Bellissent87,
  PhysRevB.76.052301} which is a cooperative analog of the Jahn-Teller
(JT) distortion\cite{Bersuker2006} in 1D or quasi-1D solids.  The
precise degree of Fermi surface nesting,\cite{CanadellWhangbo1991}
requisite for such a structural instability, seems however to be
subject to a specific approximation employed.\cite{SeoHoffmann1999,
  PhysRevB.76.052301} The special significance of
near-\emph{octahedral} coordination and the resulting trans-influence
between the corresponding covalent and secondary bonds, with regard to
the Peierls metal-insulator transition, can be viewed from yet another
angle: Alcock\cite{Alcock1972} points out that compounds in which the
two bond types display similar length show noticeable metallic luster.


We thus conclude that one should generally regard the symmetry
breaking of periodic sc parent structures as a \emph{second}-order (or
pseudo) Jahn-Teller distortion,\cite{Bersuker2006} whereby strict
electronic degeneracy is not required.
We further note that the solid-state analogs of the second-order
Jahn-Teller (JT) effect are also well known, such as the dimerization
transition in a \emph{hetero}polymer,\cite{PhysRevLett.49.1455} or in
coupled homopolar chains, such as polyacene.\cite{Burdett1995} Here,
the polymer chain is unstable toward dimerization so long as the gap
is smaller than the coupling to the symmetry breaking perturbation,
see Eq. (\ref{enegConstr}). Electronic interactions, too, can lead to
an effective one-particle gap.\cite{ZL_JCP} Although second-order JT
symmetry breaking, whenever present, is partially hampered by the lack
of degeneracy, it is still driven by the very same mechanism as during
a strict Jahn-Teller-Peierls distortion. 

We note that the question of the mechanism of the symmetry breaking is
generally distinct from that of the interaction that determines the
relaxed structure upon the symmetry breaking. According to Section
\ref{pbond}, the latter interaction in distorted-octahedral
coordinated compounds is dominated by $sp$-mixing. Indeed, Seo and
Hoffmann point out\cite{SeoHoffmann1999} that the distortion of the
relaxed structures away from the simple cubic structure is stronger
for lighter elements, consistent with stronger $sp^3$ hybridization in
those elements. For the perturbation caused by $sp$-mixing to also
contribute to the symmetry breaking itself, it is essential that this
perturbation can be made periodic with the inverse period commensurate
with the electron-filling fraction in the undistorted structure, as is
the case ($1/2=1/2$) for trans-polyacetylene and arsenic. Otherwise,
the Peierls-driven destabilization is severely weakened
(Ref.\cite{Burdett1995}, Chapter 2.6).

\begin{figure}[t]
  \centering
  \includegraphics[width=0.95\figurewidth]{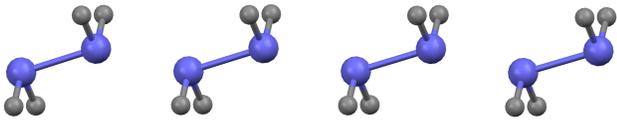}
  \caption{ \label{ASH2chain} The Peierls distorted (AsH$_2$)$_n$
    chain. The links and gaps correspond to covalent bond (bond order
    0.9) and secondary bonds (bond order 0.1) respectively.}
\end{figure}

Specifically \emph{three}-coordinated lattices with right angles
between bonds have a very special property that make them additionally
unstable toward symmetry lowering, even if these lattices are
\emph{aperiodic}.  Any such lattice\cite{burdett5764} can be thought
of as made of linear chains, \emph{each} of which consists of white
and black segments of equal length in strict alternation. The
junctions between adjacent segments correspond to the lattice
vertices, while the white and black segments themselves correspond to
no-link and link respectively.  In the case of strict octahedral
coordination, i.e., no mixing between distinct chains, \emph{each}
such chain can be thought of as a result of a Peierls distortion of a
chain of equidistant atoms. As a result, even aperiodic parent
structures are expected to be relatively (meta-)stable, not only the
strictly periodic structures of arsenic or black phosphorus.  We
illustrate the symmetry breaking in an individual linear $pp\sigma$
system in the presence of significant $sp$-mixing, which will be also
of use later.  Shown in Fig.~\ref{ASH2chain} is a dimerized linear
chain of hydrogen-passivated arsenics, (AsH$_2$)$_n$.  The details of
the electronic structure and geometry determination for the chain are
provided in Appendix \ref{app.chain}.  Despite the presence of
$sp$-mixing and other interactions, the chain exhibits a clear
$pp\sigma$-character: (see Fig.  \ref{fig_coo} for the notations) The
As-As bond length are $d=2.48$\,\AA ~and $d'=3.0$\,\AA ~for the
covalent and secondary bonds respectively. The bond angles are
$\alpha_\text{HAsH}=97^\circ$, $\beta_\text{AsAsAs}=150^\circ$.  The
secondary bonding is significant as witnessed by the value of the
corresponding transfer integral: $t' \simeq 2.3$ eV, i.e. almost a
half of that for the covalent counterpart: $t \simeq 4.9$ eV.  The
competing interactions, i.e., $ss$, $sp$-mixing, and $pp\pi$ are
significantly smaller (see Appendix \ref{app.chain}), implying the
bonding is indeed $pp\sigma$.  The $sp$-transfer integral is as large
as 57\% of the $pp\sigma$ transfer integral, consistent with the
earlier statement that the $sp$-mixing is the next leading contributor
to the geometry of the symmetry broken state, after the $pp\sigma$
interaction itself. As expected of a Peierls insulator, the chain
exhibits a perfect bond alternation pattern.  Lastly note that even
though individual chains that make up aperiodic 3D parent structures
undergo Peierls transitions, the symmetry breaking for the 3D
structures themselves is generally not quasi-one dimensional and thus
is not Peierls, in contrast with periodic systems, such as considered
by Burdett and others.

The parent structures with three links per atom plus vacancies, if
any, see Figs.~\ref{Burdett}-\ref{As2Se3Xtal}, can serve as basic
models for compounds with distorted octahedral coordination that
consist of two- and three-valent elements. These structures also apply
to compounds containing elements from groups 14, such as GeTe, GeSe
from Fig.~\ref{GeTeXtal} or archetypal phase-change alloys
Ge$_2$Sb$_2$Te$_5$ and GeSb$_2$Te$_4$.\cite{PhysRevB.77.035202} Hereby
each A$^\text{IV}$B$^\text{VI}$ pair is isolectronic with a pair of
pnictogens, implying these atoms are three-coordinated. The rest of
the atoms are in the Pn$_2$Ch$_3$ stoichiometry, and so the rules
leading to the parent structure in Fig.~\ref{As2Se3Xtal} apply to
these atoms. Even in the absence of A$^\text{IV}$B$^\text{VI}$
pairing, symmetry breaking schemes can be proposed for substances
where the number of four-coordinated atoms is large, as in GeP and
TlI,\cite{burdett1434} or $\beta$-tin,\cite{BurdettLeePeierls} and
similarly for five-coordinated atoms.\cite{BurdettLeePeierls}
Otherwise, an atom with coordination 4 or higher can be considered as
a defect in a lattice of three-coordinated vertices. We will return to
this important point in the companion article.\cite{ZLMicro2}

The following picture of structure-formation in $pp\sigma$-networked
materials thus emerges from the above considerations: These materials
may be thought of as symmetry-broken versions of a simple cubic
structure. The symmetry breaking is an interplay of several types of
perturbation: (a) Peierls-instability proper for extended near-linear
chains; (b) cooperative second-order Jahn-Teller distortion that
results from mixing of the $p$-orbitals with other orbitals, mostly
$s$, and from electronic interactions; (c) steric effects due to
vacancies, if the latter must be present owing to stoichiometry, as in
Fig.~\ref{As2Se3Xtal}; and (d) other coordination variations, as in the
case of elements from groups 14 and lower.  Even though these
perturbations are strong enough to break the symmetry, they are still
\emph{perturbations}, so that the resulting bonding is primarily
$pp\sigma$. This statement can be quantified by comparing the
strengths of the corresponding transfer integrals in the deformed
structure, see Table~\ref{tab.tint} and the (AsH$_2$)$_n$ example
above.

Now, one must recognize that upon geometric optimization, the lattice
will generally be \emph{aperiodic}, even when the link-breaking
pattern in the parent structure is itself strictly periodic, let alone
if we arranged the distinct species or vacancies aperiodically or with
a period incommensurate with the period prescribed by the electron
filling fraction.  Let us now examine how such aperiodic lattices
maintain the $pp\sigma$ character, and hence the stability with
respect to other types of ordering, while, at the same time, allowing
for molecular transport.

For concreteness, let us consider a specific prescription to deform
the parent sc structure. For each atom, consider the links in the
parent structure as \emph{vectors} that start on the atom.  Move each
atom by a small distance in the direction which is the sum of its own
vectors. For instance, if an atom has three links: (1, 0, 0), (0, -1,
0), and (0, 0, -1), move it in the direction (1, -1, -1). Analogously
for a two coordinated atom, the displacement will be in the plane
containing the two links. Now turn on the interaction in the form of
non-zero transfer integrals, such as listed in Tables \ref{tab.tint}
and \ref{tab.coo}. Let the lattice relax, subject of course to the
Coulomb repulsion between the ionic cores. As already mentioned, there
is generally no one-to-one correspondence between the parent and the
distorted structure: this implies there are multiple relaxed
structures and hence \emph{multiple metastable minima on the total
  energy surface}. In discussing
Figs.~\ref{As2Se3Xtal}-\ref{linksSide}, we have mentioned that
distinct parent structures for the Pn$_2$Ch$_3$ stoichiometry can be
obtained by shifting the double layers relative to each other. Because
such a shift incurs bond breaking, these alternative parent structures
are separated by barriers. It is understood that since the parent
structure is generally aperiodic,\cite{ZLMicro2}
Figs.~\ref{As2Se3Xtal}-\ref{linksSide} apply only to a small fragment
of such an aperiodic structure. Now, since the distinct parent
structures are separated by an energy barrier, at least one of them
should be separated by a barrier from the actual deformed structure,
implying a presence of additional, metastable minima.  When such
metastable minima are few, the global potential energy minimum is
easily accessible, as is probably the case for the periodic parent
structure of arsenic, see Fig.~\ref{Burdett}(a). Elemental arsenic is,
in fact, a poor glassformer.  (Other distinct three-coordinated parent
structures exist,\cite{burdett5764} but most of them are energetically
costly.\cite{burdett5774}) To summarize, the existence of distinct
parent structures with shifted atoms is crucial for the present
structural model to be consistent with the presence of molecular
transport in $pp\sigma$-networks.

When aperiodic, the distorted lattice will exhibit two key features:
First, if the substance forms a periodic crystal, it will be lower in
energy than any aperiodic counterpart. Second, because aperiodic
structures are not unique, the lattice will be highly degenerate, as
just discussed. We can also understand the emergence of the degeneracy
thermodynamically: Suppose the substance can crystallize. A
mechanically stable aperiodic structure, on the one hand, has a much
lower symmetry than the crystal. On the other hand, the aperiodic
structure corresponds to a higher energy and hence higher
\emph{temperature}. The only way to reconcile these two conflicting
notions is to recognize that there should be a thermodynamically large
number of nearly degenerate aperiodic structures separated by finite
barriers. By virtue of barrier crossing events, the lattice is able to
restore the full translational symmetry at long enough times; the
latter symmetry is higher than that in a mechanically stable
crystal. The lattice therefore corresponds to a \emph{liquid} in the
activated transport regime, if steady state,\cite{LW_soft} or to an
aging glass, if below the glass transition.\cite{LW_aging}

The view of quenched liquids and frozen glasses as broken symmetry
phases is supported by an independent argument: According to
Fig.~\ref{symmetrybreaking}, the symmetry broken phase corresponds to
a lower pressure. Bevzenko and Lubchenko\cite{BL_6Spin} have shown
that a covalently-bonded equilibrium melt can be regarded as a high
symmetry lattice that has been sufficiently \emph{dilated} and then
allowed to relax into one of the many available aperiodic
configurations. Now, are the predicted structural degeneracy of
emergent aperiodic $pp\sigma$-networks and the barriers for activated
reconfigurations consistent with the configurational entropy of actual
materials? The contiguity between covalent and secondary bonding, as
illustrated in Fig.~\ref{symmetrybreaking}, suggests that
$pp\sigma$-networks support atomic motions that do not involve
bond-breaking but only a gradual change in the coordination, and as
such, may be thermally accessible. These special atomic motions will
be discussed in detail in the follow-up article.\cite{ZLMicro2}

Available structural data are consistent with the prevalence of
$pp\sigma$-bonding in \emph{vitreous} chalcogenides with the
stoichiometries in question. According to several
studies,\cite{Iwadate19991447, Hosokawa99} the nearest neighbor bond
lengths in amorphous arsenic chalcogenides are essentially identical
on the average to those in their crystalline counterparts, but have a
somewhat broader distribution. In view of the argued presence of
trans-influence between the covalent and secondary bonds in these
compounds, we conclude that the secondary bonds in the vitreous
samples are of strength comparable to those in the corresponding
crystals, again subject to a broader distribution.  Consequently,
based on the applicability of the same general mechanism of
$pp\sigma$-network formation and the comparable bonding strength, we
conclude both periodic and aperiodic lattices exhibit the same type of
bonding.  Still, for the sake of the argument, suppose on the contrary
that the bonding is dominated not by the $pp\sigma$ interaction, but
by its leading competitor, i.e., $sp$-mixing, thus resulting in a
predominantly tetrahedral coordination. The ratio of the filling
fractions of the diamond and As$_2$Se$_3$ lattices is approximately
1.18. At the same time, the densities\cite{Iwadate19991447,
  Stergiou85, Madelung} of the amorphous and crystalline compounds
differ by less, i.e.  is 4.57 vs. 4.81-5.01 g/cm$^3$ for As$_2$Se$_3$
and 3.19 vs. 3.46 g/cm$^3$ for As$_2$S$_3$, implying that only a small
fraction of atoms in these glasses, if any, might be regarded as
tetrahedrally coordinated. Such a ``defect'' is analogous to a small
region occupied by an interface between two distinct lattices, which
incurs a significant free energy cost. The companion
article\cite{ZLMicro2} shows that a mechanism for a reversible
formation of such defected configurations arises naturally in the
present structural model.

\section{Concluding Remarks}

The main goal of this article was to establish the mechanism of
bonding in semiconducting pnictogen- and chalcogen-containing quenched
melts and frozen glasses. Representative substances are archetypal
glassformers, such as As$_2$Se$_3$ and similar materials whose
crystalline forms can be directly argued to exhibit
$pp\sigma$-bonding, based on their known structures, measured
electronic density of states, and electronic structure calculations.
We have formulated a structural model, by which both the crystalline
and vitreous materials are seen to form by the same general mechanism,
i.e., by symmetry-lowering and distortion of a high-symmetry structure
defined on the simple cubic lattice. By combining this model with the
limited structural data on the vitreous counterparts of those listed
materials and similar compounds, we have argued the glasses are also
$pp\sigma$-bonded.

Lowering of the symmetry by breaking the bonds in the fully connected
simple cubic structure can be understood as lowering of the lattice
dimensionality, similarly to what is seen in the (AsH$_2$)$_n$ chain
in Fig.~\ref{ASH2chain}, which is a linear array of relatively weakly
bonded dimers, or to what one finds in the parent structures from
Figs.~\ref{Burdett}-\ref{As2Se3Xtal}, which consist of double-layers,
possibly accompanied by further symmetry lowering. Papoian and
Hoffmann\cite{PapoianHoffmann2000} have outlined general principles
for the interrelation of dimensionality and deformation of high
symmetry periodic structures. These authors argue the bonding in the
parent structures from Fig.~\ref{Burdett} can be regarded as a Peierls
distortion of a hypervalently bonded lattice built from electron-rich
units, whereby the dimensionality of the lattice is lowered. In
contrast, to \emph{preserve} the dimensionality, the electron-rich
units comprising the hypervalent structure would have to be oxidized
instead. In the former case, the deformed structure retains its
original character, while in the latter case, the geometry is expected
to change. For instance, a cubic Sb$^0$ lattice exhibits a (distorted)
octahedral coordination, while oxidation would hypothetically result
in a tetrahedrally coordinated Sb$^+$ lattice. The semiconducting
alloys in question do exhibit relatively low variation in
electronegativity and, hence, the distorted octahedral coordination.

The parent structures for substances at the focus of the present
study, i.e., pnictogen and chalcogen containing alloys enjoy a very
special property: Each vertex in the parent structure is
\emph{three}-coordinated, while the bonds are at near $90^\circ$
angles. Under these special circumstances, the whole lattice can be
thought of as composed of infinite linear chains in which bond and
bond-gaps strictly alternate. In view of the weak interaction with the
environment, each of these chains can be thought of as a quasi-1D
Peierls insulator with renormalized interactions. (In the eventual
deformed structure, the chains are deformed and, likely, of rather
limited length.\cite{ZLMicro2}) This observation allows one to extend
the trends established by Papoian and Hoffmann in periodic crystals to
\emph{aperiodic} lattices. Since the symmetry can be restored
partially or fully by high pressure, the argued view of glasses as
lowered-symmetry versions of high-symmetry structures is supported by
independent arguments on (negative) pressure-driven glass transition
in covalently bonded materials.\cite{BL_6Spin} We have also pointed
out that the view of this type of symmetry breaking as a second-order,
cooperative Jahn-Teller distortion may be equally justified.  In
addition to the Peierls instability proper, within each chain, the
symmetry breaking is also driven by local interactions that compete
with the $pp\sigma$-bonding proper, primarily by the $sp$-mixing.

The presence of competing interactions in these glass-forming
materials is consistent with their structural
degeneracy.\cite{Lcompeting} It appears that the strength of such
competing interactions should satisfy certain
restrictions. Specifically, if the $sp$-mixing is too strong, it
destroys the $pp\sigma$-bonding. Yet, if the mixing is sufficiently
weak, the coordination can be made close to perfect octahedral, while
decreasing the stability of the glass relative to the crystal.  The
latter trend is exploited in making optical drives, using
Ge$_2$Sb$_2$Te$_5$ or similar compounds.\cite{Kolobov2004,
  Steimer2008} Again consistent with the structural degeneracy of the
$pp\sigma$-networks is the noted analogy between the bond-placement
rules in the proposed structural model and the vertex models known to
exhibit rich phase behavior.\cite{Baxter, 714506920011010}

The argued similarity of the bonding mechanisms in $pp\sigma$-bonded
crystals and glasses explains the puzzling stability of this important
class of glasses. As mentioned in the Introduction, the enthalpy
excess $\Delta H$ of the supercooled liquid, relative to the
corresponding crystal, is easy to estimate. It is directly related to
the configurational entropy $S_c$ of the fluid, i.e. $\Delta H = S_c
T$, save a small ambiguity stemming from possible differences in the
vibrational entropies. The liquid configurational entropy varies
between $0.8 k_B$ and about $2k_B$ per bead, between the glass
transition\cite{XW} and melting temperatures. The energy $0.8 k_B T_g$
per bead amounts to less than 0.05 eV per atom. The model thus allows
one to reconcile two seemingly conflicting characteristics of quenched
melts and frozen glasses. On the one hand, these materials exhibit
remarkable thermodynamic and mechanical stability, only slightly
inferior to the corresponding crystals. On the other hand, these
materials are also multiply degenerate thus allowing for molecular
transport.

Finally, this paper buttresses and clarifies some of the technical
aspects of the original programme: The $pp\sigma$-networks
automatically satisfy all the necessary requirements for the presence
of the intrinsic midgap electronic states, as listed in the
Introduction. First, a necessary requirement for covalent $pp\sigma$
bonding is that the electronegativity variation $\epsilon$ is not too
large, Eq. (\ref{enegConstr}), lest the resulting bonding becomes
predominantly ionic. Second, a high-symmetry $pp\sigma$-bonded network
is unstable toward Jahn-Teller distortion at each center, implying the
deformed lattice shows an alternating pattern of bond saturation in
the form of covalent and secondary bonds.  Third, based on the
stability of the $pp\sigma$-network with respect to perturbations in
the form of competing chemical interaction, the secondary bonds are
sufficiently strong, i.e. the $t'/t$ ratio of the transfer integral of
the secondary and covalent bonds is not too small, see
Eq.~(\ref{tratio}).

{\it Acknowledgments}: The authors thank David M. Hoffman, Thomas
A. Albright, Peter G. Wolynes, and the anonymous Reviewer for helpful
suggestions.  We gratefully acknowledge the Arnold and Mabel Beckman
Foundation Beckman Young Investigator Award, the Donors of the
American Chemical Society Petroleum Research Fund, and NSF grant
CHE-0956127 for partial support of this research.

\appendix

\section{Detailed explanations of Tables \ref{tab.tint} and
  \ref{tab.coo}} \label{app.tint}

\begin{figure}[t]
\centering
\includegraphics[width= 0.9 \figurewidth]{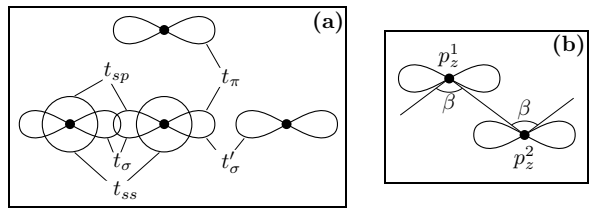}
\caption{(a) A graphical summary of the definitions of the transfer
  integrals from Table~\ref{tab.tint}. (b) An example of geometry used
  to compute transfer integrals $t$ and $t'$, see text.}
\label{tdef}
\end{figure}

Table~\ref{tab.tint}: A graphical summary of the definitions of the
transfer integrals is given in Fig.~\ref{tdef}(a), see also p. 23 of
Ref.\cite{Harrison} By definition, the $pp\sigma$ integrals, i.e.,
$t_\fn$ and $t_\fn'$ are computed assuming the two $p$ orbitals are
aligned with the bond and hence depend only on the distance between
the participating atoms. As a result, the transfer integrals $t$ and
$t'$ from Eq.~\ref{tratio} are equal in value to $t_\fn$ and $t_\fn'$
respectively only when the three atoms in question are situated on a
straight line. If $180-\beta<20^\circ$, as is the case for almost all
compounds from Table~\ref{tab.tint}, the difference between $t$ and
$t_\sigma$ is less than a few percent, depending on the specific
geometry and basis set. For instance, in the geometry from
Fig.~\ref{tdef}(b), we get $t=\left\langle p_z^1 \right| \cH \left|
  p_z^2\right\rangle=t_\sigma+(t_\pi-t_\sigma)\cos^2(\beta/2)$,
yielding a difference of 4\% between $t$ and $t_\sigma$ for $\beta =
160^\circ$ and $t_\pi/t_\sigma = - 1/4$, Ref.~\cite{Harrison04}

Table~\ref{tab.coo}: For several compounds, the structural data vary
somewhat depending on the source, such as for As$_4$Se$_4$ in
Refs. \cite{Renninger73, Goldstein74} and $\alpha$-monoclinic Se,
Refs. \cite{Wyckoff63, Cherin72} The resulting ambiguity should be
kept in mind.

\emph{2nd column}, deviation from the right-angled geometry. For
As$_2$O$_3$, GeSe$_2$, and AsBr$_3$ only As and Ge were considered as
the central atoms.

\emph{3rd column}. The absolute electronegativity is the average of
the electron ionization and affinity energies, as found in
Ref.\cite{NIST}

\emph{4th, 5th column}. The covalent radii used are listed in Table
\ref{tab.rcov}. These radii were determined using compounds exhibiting
low variations in electronegativity, as pertinent to the materials in
question, except GeO$_2$.  For As, values found in the
literature\cite{Cordero08, Pyykko09, CCDC} differ within 0.02 \AA. We
use the value from the middle of this range, which also happens to
coincide with the result of interpolation across the sequence of
Ge-As-Se-Br.

\begin{table}[t]
  \centering
  \begin{tabular}{|c|l|c|} \hline
    & reference structure & $r_\text{cov}$,\AA \\ \hline
    O  & $\alpha$-quartz-GeO$_2$ crystal & 0.52 \\
    Ge & crystal & 1.225 \\
    As & see text & 1.20 \\
    Se & isolated helix \cite{Springborg09} and rings & 1.17 \\
    Br & diatomic molecule & 1.140 \\
    Te & isolated helix and rings \cite{Ghosh08} & 1.37 \\ \hline
  \end{tabular}
  \caption{Covalent radii. The three columns contain the species name, 
    the reference structure, and the resulting radius.}
  \label{tab.rcov}
\end{table}

\emph{6th column}. Only the strongest secondary bonds are cited. To
determine these for As$_4$Se$_4$ and Se, we have used diagrams that
show the magnitudes of the atomic $p$ orbitals of the nearby atoms on
a sphere centered on a chosen atom with a radius equal to the covalent
radius of that atom, see the Supplementary Material.\cite{SupplMat}
For the tetrahedrally bonded materials, As$_2$O$_3$, and AsBr$_3$, the
nearest neighbor in the direction opposite to the covalent bonded
atoms is used.

\emph{7th column}. Accurate values of the tight-binding (TB)
parameters are usually determined by obtaining the best fit to the
electronic density of states for each specific systems and are thus
system-dependent.  Our goal here is, instead, to highlight the
\emph{generic} trends of tight binding parameters that apply
satisfactorily for bonds ranging from the strictly covalent to van der
Waals, in as many distinct compounds as possible. Such a universal
parametrization of one-electron transfer (resonance) integrals is
provided, for instance, by the PM6 parametrization \cite{Stewart07} of
the MNDO method,\cite{Dewar77} which we use to estimate the
$t'_\sn/t_\fn$ ratios for the compounds in Table~\ref{tab.coo}.
Despite the made approximations, we feel that the resulting potential
ambiguity of the \emph{ratio} of the matrix elements, $t'_\sn/t_\fn$,
is not large.

We can partially judge the reliability of the tight-binding parameters
by comparing them to those obtained by other standard methods.  In
Table~\ref{tab.TBparam} below, we compare Robertson's
data\cite{Robertson83} for transfer integrals in three elemental
solids, as obtained by fitting the spectrum of the one-electron TB
Hamiltonian (column 2) and by the chemical pseudopotential method
(column 3) to the present TB parametrization (column 4). The Figure on
the r.h.s. compares the result of the parametrization for elemental Ge
with accurate calculations of Bernstein et al.\cite{Bernstein02}
Finally, the same TB parametrization was used for calculation on the
dimer of Br$_2$ molecules, see below.
\begin{table}[t]
  \centering
  \begin{minipage}[l]{0.45\figurewidth}
    \centering
    \begin{tabular}{|c|c|c|c|} \hline
      &  Ref.\cite{Robertson83} & Ref.\cite{Robertson83} & present \\
      & TB fit & pseud. & work \\ \hline
      As & 0.54 & 0.51 & 0.62 \\
      Se & 0.19 & 0.26 & 0.27 \\
      Te & 0.33 & 0.65 & 0.74 \\ \hline
    \end{tabular}
  \end{minipage} \hspace{3mm}
  \begin{minipage}[r]{0.45\figurewidth}
    \centering
    \includegraphics[width=\textwidth]{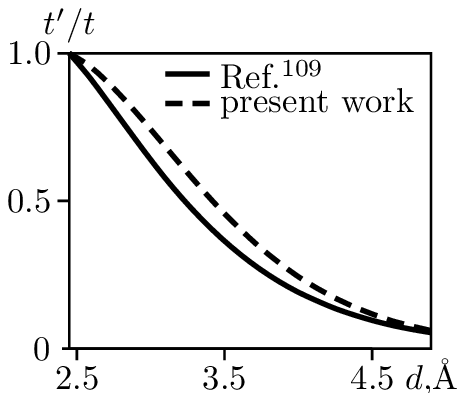}
  \end{minipage}
  \caption{Comparison of the $t'_\sn/t_\fn$ ratio, as resulting from 
    the present tight-binding parametrization with compilations  of
    Robertson\cite{Robertson83} (table) and the calculations of Bernstein 
    et al.\cite{Bernstein02}  (figure), see text.}
  \label{tab.TBparam}
\end{table}

\section{Dimer of bromine molecules} \label{app.dimer}

In this appendix we use the electronic structure of Br$_2$--Br$_2$
dimer obtained by ab-initio calculations to estimate the contribution
of the $pp\sigma$ interaction to the binding energy of the dimer and
show that the resulting estimate is consistent with the conclusions of
a simple molecular orbital theory.

The geometry of the dimer (see Fig.~\ref{fig_Brdimer}) is optimized by
Firefly \cite{Firefly} program on MP2-level with aug-cc-pVTZ basis set
and RHF wave-function. Although more accurate approximations
exist,\cite{Moilanen09} the present method has the advantage of
simplicity while yielding the geometry consistent with that of the
bromine crystal; it also yields results that compare well with
accurate calculations for other halogen dimers.\cite{Karimi-Jafari09}
For clarity, we have not corrected for the basis set superposition
error, because here we are interested in the relative magnitudes of
distinct contributions to the bonding, not the absolute value of the
binding energy. The resulting ambiguity in the binding energy itself
is not large anyway, about 15\%, as can be estimated using the
standard counterpoise method, Ref.\cite{Levine09} p. 714.

The total binding energy of the dimer, 0.13~eV, consists of several
contributions.  For instance, the correlation (MP2) contribution is
0.2~eV. To extract the contribution proper of the electrons occupying
distinct molecular orbitals, we adhere to the following scheme: First,
we subtract the uniform downshift of all MOs by 20~meV in the dimer,
relative to the two isolated Br$_2$ molecules. Next, we compile the
contributions of all molecular orbitals to the bonding, see
Table~\ref{tab.dimerMO}. Here, the dimer is in the x-y plane. Because
little $sp$-mixing is present, the MO's are naturally grouped into
classes that consist primarily of either $p$ or $s$ orbitals. The $p$
orbitals are further subdivided, according to their symmetry, into
$p_{x,y}$ and $p_z$ orbitals.  The former form the $pp\sigma$ bonds,
while the latter are out of the x-y plane and contribute little to the
bonding, as is clear from Table~\ref{tab.dimerMO}.

\begin{table}[h]
  \begin{equation}
    \begin{array}{|c|c|c|r|} \hline
      \text{MOs} & \text{AOs} & \text{energy range, eV} & E_\text{binding}, \text{eV} \\ \hline
      61,62,64,66,68,69      & p_{x,y}   & (-14.90,-10.9) & -0.284 \\ 
      57,58,59,60            & s         & (-30.3, -25.7) & -0.195 \\
      63,65,67,70            & p_z       & (-14.0,-10.9) & -0.075 \\ \hline
      71,72 \text{ (unocc.)} & s,p_{x,y} &  (-0.3, +0.2) & +0.466 \\ \hline  
    \end{array} \nonumber
  \end{equation}
  \caption{ 
    Bonding contribution of the molecular orbitals (MOs) of the Br$_2$-Br$_2$ 
    dimer from Fig.~\ref{fig_Brdimer}.   The quantity
    $E_\text{binding}$ is the difference between the total MO energies
    of the dimer and those of the isolated molecules.
  }
  \label{tab.dimerMO}
\end{table}

\begin{figure}[t]
\centering
\includegraphics[width= 0.9 \figurewidth]{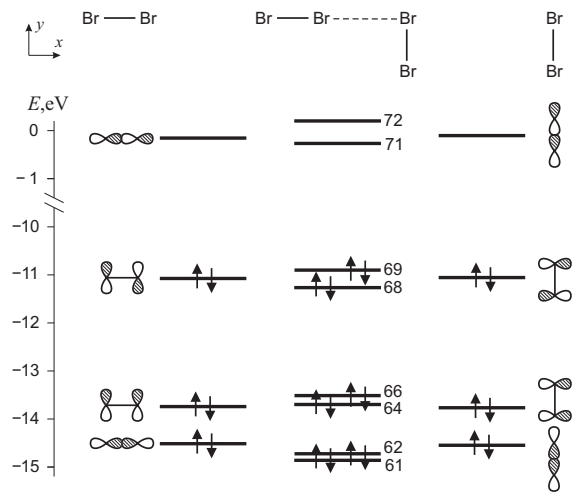}
\caption{MO diagram of the Br$_2$ dimer from Fig.~\ref{fig_Brdimer}
  that depicts a subset of the MO's from Table~\ref{tab.dimerMO},
  along with pertinent MO's of the constituent Br$_2$ molecules.}
\label{BrMO}
\end{figure}

The magnitude of the binding energy due to the in-plane $p$-orbitals,
0.284~eV, can be rationalized using a simple molecular orbital
theory. Let us first consider only the eight valence $p$-orbitals
lying in the plane of the dimer shown in Fig.~\ref{fig_Brdimer}. The
{\it ab initio} diagram of the corresponding MO's is shown in
Fig.~\ref{BrMO}. Let us focus exclusively on the intramolecular
$pp\sigma$- and $pp\pi$-interactions and the intermolecular
$pp\sigma$-interaction between two atoms connected by the dashed line
in Fig.~\ref{fig_Brdimer}. In this approximation, the $p_y$ orbitals
do not interact with the $p_x$ orbitals, so that the secondary bonding
is exclusively due to the $p_x$ orbitals.  The specific realization of
Hamiltonian (\ref{Hab}) for this bonding configuration reads
\begin{equation} \label{Hpx} \cH=\begin{pmatrix} \epsilon_p & t_\fn &
    0 & 0 \\ t_\fn & \epsilon_p & t'_\sn & 0 \\ 0 & t'_\sn &
    \epsilon_p & t_\pi \\ 0 & 0 & t_\pi & \epsilon_p \end{pmatrix}.
\end{equation}
By Eq.~(\ref{Ebind}), combined with
\begin{gather*}
  E^\text{A}_\text{unocc}=\epsilon_p+t_\fn, \qquad
  \psi^\text{A}_\text{unocc}=
  \frac{1}{\sqrt{2}}\begin{pmatrix} 1 \\ 1 \end{pmatrix}, \\
  E^\text{B}_\text{occ}=\epsilon_p\pm t_\pi, \qquad
  \psi^\text{B}_\text{occ}=\frac{1}{\sqrt{2}}\begin{pmatrix}
    1 \\ \pm 1 \end{pmatrix}, 
\end{gather*}
we obtain the dimer binding energy:
\begin{equation} \label{dimerE} (t'_\sn)^2/t_\fn.
\end{equation}
Generally, in the perturbative expression (\ref{Ebind}), the
quantities $V_{nm}$ correspond to the transfer integrals of the
secondary bonds, $t'$, while the denominators to the transfer
integrals of the covalent bonds, $t$. The binding energy is thus
second order in the $t'/t$ ratio.

Note that the result in Eq.~(\ref{dimerE}) also helps to partially
assess the reliability of the TB parametrization from
Appendix~\ref{app.tint}. Substituting the numerical values for those
transfer integrals in Eq.~(\ref{dimerE}) yields the dimer bonding
energy 0.3~eV, in good agreement with the ab-initio analysis leading
to Table~\ref{tab.dimerMO}. A similar analysis can be used to show
that the $s$-orbital bonding contribution from Table~\ref{tab.dimerMO}
results not from the $ss$ interaction, but primarily from $sp$-mixing,
also consistent with the small overlap of the $s$-orbitals.

Finally, we outline derivation of the TB expression (\ref{BondOrder})
for the bond order of a closed-shell interaction.  Using Eq.~(14.22)
of Ref.\cite{MurrellKettleTedder}, for the wave-functions of an AB
dimer formed by a closed shell interaction, one can straightforwardly
show that the interaction-induced correction to the density matrix
${\cal P}$ for the dimer, in the second order, is given by

{\footnotesize
  \begin{equation}
    \left\{\sum_{n}^{\text{A$_\text{occ}$}}\sum_{m}^{\text{B$_\text{unocc}$}}
      -\sum_{n}^{\text{A$_\text{unocc}$}}\sum_{m}^{\text{B$_\text{occ}$}}\right\}
    \frac{V_{nm}}{E_n^\text{A}-E_m^\text{B}}
    \begin{pmatrix} 0 & |\psi_n^\text{A}\rangle\langle\psi_m^\text{B}| \\
      |\psi_m^\text{B}\rangle\langle\psi_n^\text{A}| & 0 \end{pmatrix}. ,
  \end{equation}}

According to a standard definition of the bond
order,\cite{Armstrong73} (also used in MOPAC), the latter can be
written in the chosen basis set as:
\begin{equation} b_\text{AB} = \sum_n^\text{A} \sum_m^\text{B} \left|(
    \la \psi_n^\text{A}|, 0)  {\cal P} 
    \begin{pmatrix}0 \\ | \psi_m^\text{B} \ra  \end{pmatrix} \right|^2,
\end{equation}
leading to Eq.~(\ref{BondOrder}).

Note that the first (second) double sum in Eq.~(\ref{BondOrder}) is
twice the occupation of the antibonding orbitals of the molecule B
(A), as donated by the bonding orbitals of A (B). These occupations
are defined as
$2\sum_{\nu}^{\text{(AB)$_\text{occ}$}}\sum_{m}^{\text{B$_\text{unocc}$}}
| \left\langle\psi_m^\text{B}\big|\psi_{\nu}\right\rangle|^2$ and
$2\sum_{\nu}^{\text{(AB)$_\text{occ}$}}\sum_{m}^{\text{A$_\text{unocc}$}}
|\left\langle\psi_m^\text{A}\big|\psi_{\nu}\right\rangle|^2$
respectively, where $\psi_{\nu}$ are MO's of the AB dimer.

\section{Arsenic chain} \label{app.chain}

In our model system, the dimerized (AsH$_2$)$_n$ chain
(Fig.~\ref{ASH2chain}), the shorter and longer As--As bonds correspond
to covalent and secondary bonds respectively; the As--H bonds
correspond to covalent bonds in chalcogenide glasses.  The
calculations are performed on semiempirical level by MOPAC
\cite{MOPAC} program with PM6 parametrization. On each arsenic, the
axes are labelled as follows: The $p_x$ and $p_y$ orbitals are
oriented toward the hydrogens (but not strictly along the x and y
axis), while $p_z$ orbitals are directed toward the neighboring
arsenics, see the geometry in Fig.~\ref{tdef}(b). Note these axes are
defined locally on each individual arsenic; the optimized chain may or
may not be linear on the average.

We first check the semiempirical approximation against ab-initio
calculations for $n=4$, see Table~\ref{tab.Asdimer}. The {\it ab
  initio} results clearly indicate a distorted octahedral geometry and
are consistent with secondary bonding between the terminal
(AsH$_2$)$_2$ units: The $\beta$ (As-As-As) angle is very close to
$180^\circ$, while the bond length, 3.57 \AA ~is close but shorter
than the sum of the reported van der Waals radii for As, i.e., between
3.7 and 3.9 \AA, depending on the convention.\cite{Bondi64} Now,
judging from the angle $\beta$ between adjacent As--As bonds and
H-As-H angle, the semiempirical method clearly overestimates the
tendency toward $sp^3$ hybridization. At the same time, the
semiempirical method underestimates the secondary bond length,
suggesting the two errors in the resulting bond strength will
partially cancel out. The discrepancy in the secondary bond lengths
and the $\beta$ angles between the {\it ab initio} and PM6 methods
should be considered a consequence of the specific parametrization
adopted in MOPAC, which was optimized for covalent, not secondary
bonds, and many of which have a significant ionic component and/or
coordinations distinct from distorted octahedral.  Thus the TB-based
analysis is not expected to be fully quantitative, unless ad hoc
parametrized. Nevertheless, we have seen from the discussion of
Eqs.~(\ref{Ebind}) and (\ref{BondOrder}) that TB analysis is
qualitatively consistent with the closed-shell character of the
secondary bonding and the trans-influence between the counterpart
covalent and secondary bonds.  Our results for geometric optimization
of rhombohedral arsenic, using the PM6 parametrization, are consistent
with these conclusions: Using a 96 atom supercell, we obtain $d =
2.46$ \AA ~(vs. experimental 2.52 \AA), $d' = 3.24$ \AA ~($3.10$ \AA),
$\alpha = 99.3^\circ$ ~($96.7^\circ$). Clearly, the PM6
parametrization is consistent with $pp\sigma$ bonding in rhombohedral
As, and, furthermore, gives reasonable figures for the lattice
parameters.

\begin{table}[t]
\centering
\begin{tabular}{|l|cccccc|c|} \hline
& $d_\text{AsAs}$ & $d'_\text{AsAs}$ & $\beta$ & $d_\text{AsH}$ & $\alpha_\text{HAsH}$ & $\alpha_\text{AsAsH}$ & $E$,eV \\ \hline
PM6 & 2.463 & 3.06 & 148 & 1.523 & 95.7 & 97.4 & -0.17 \\
                     &&&& 1.523 & 95.6 & 96.9 & \\ \hline
MP2 & 2.483 & 3.57 & 177 & 1.525 & 91.7 & 92.6 & -0.13 \\
                     &&&& 1.527 & 91.6 & 92.1 & \\ \hline
Ref.\cite{Klinkhammer95} & (2.441) & 3.53 & (180) & (1.506) & (93.9) & (93.9) & -0.10 \\ \hline
\end{tabular}
\caption{The geometry of the dimer of As$_2$H$_4$ molecules: semiempirical PM6 calculations, ab-initio RHF MP2 calculations with acc-pVDZ basis set, and calculations from Ref.\cite{Klinkhammer95} The notations are according to Fig.~\ref{fig_coo}, two entries per cell correspond to the inner (1st) and the outer (2nd) AsH$_2$ units, the energy is the dimer binding energy, the values in parentheses are not optimized.}
\label{tab.Asdimer}
\end{table}

According to the semi-empirical calculation, an \emph{infinite}
AsH$_2$ chain is dimerized and has a zero net curvature, whereby
$d=2.48$\,\AA, $d'=3.0$\,\AA, $\alpha=97^\circ$, $\beta=150^\circ$.
Despite the aforementioned bias toward $sp$-mixing, analysis of
localized MOs in the semiempirical calculations shows that the bonding
contribution of arsenic's $s$-orbitals is negligible: 0.1 (vs. 2 for a
regular bond).  Significant secondary bonding and the trans-influence
between the covalent secondary bonds are clearly seen in the values of
the bond order, which are equal to 0.9 and 0.1 for the covalent and
secondary As-As bond respectively.

\begin{table}[t]
\centering
$$\begin{array}{|l|rrrr|c|c|} \cline{2-7}
  \multicolumn{1}{c|}{\phi\,\backslash\,\psi}
  &  p_z &   s  &  p_x & s_\text{H} & \text{\scriptsize renormalization} &  sp  \\ \hline
  p_z            & -4.9 & -0.6 &  0.2 &  0.1 & \tilde{\epsilon}  - \epsilon = +1.2 \text{eV} & 60\%  \\
  s              &      & -13. &  0.9 & -4.0 && \\
  p_x            &      &      & -4.4 & -7.6 && \\
  s_\text{H}     &      &      &      & -5.0 && \\ \hline
  p_z^\text{cb}  &  4.9 (t) &  2.8 & -0.8 &  0.6 & \tilde{t}  - t =  +0.7 \text{eV} & 50\%  \\
  s^\text{cb}    &      & -0.7 &  0.4 & -0.1 && \\
  p_x^\text{cb}  &      &      &  1.3 &  0.4 && \\ \hline
  p_z^\text{sb}  &  2.3 (t') & -1.0 &  0.7 & -0.2 & \tilde{t}' - t' = -0.3 \text{eV} & 35\%  \\
  s^\text{sb}    &      &  0.0 &  0.2 & -0.1 && \\
  p_x^\text{sb}  &      &      &  0.5 &  0.2 && \\ \hline
\end{array}$$
\caption{
  Columns 2-5: Elements $\langle\phi|H|\psi\rangle$ of the Fock matrix
  of an infinite (AsH$_2$)$_n$ chain. The transfer integrals $t$ and $t'$ 
  are indicated in brackets next to their numerical values. Column 6: The renormalization of
  the parameters of the one-particle $pp\sigma$-only Hamiltonian, as
  computed using Eq. (\ref{renorm}). Column 7: The contribution to this
  renormalization of the $s$-orbitals. Labels ``cb'' and ``sb'' mean
  covalent and secondary bonds. All AO's belong to the arsenics except the 
  hydrogen's $s_\text{H}$.
}
\label{tab.FockM}
\end{table}

To independently verify the dominant character of the
$pp\sigma$-bonding, we demonstrate that the competing interactions are
perturbations that mostly amount to a renormalization of the
$pp\sigma$-interaction.  Indeed, the full \emph{one}-particle
Hamiltonian for a system plus the environment is a block matrix
\begin{equation} \label{block} \cH = \left( \begin{array}{cc}
      \cH_\text{sys} & \cV^\dagger \\ \cV & \cH_\text{env}
    \end{array} \right),
\end{equation}
where the matrix $\cH_\text{sys}$ contains exclusively the on-site
energies and transfer integrals for the system, $\cH_\text{env}$ for
the environment. The (generally non-square) matrix $\cV$ and its
hermitian conjugate $\cV^\dagger$ contain the system-environment
transfer integrals. It is straightforward to show that, given an
energy eigenvalue $E$ for the full Hamiltonian in Eq.~(\ref{block}),
the portion of the wave-function corresponding to the system is an
eigenfunction of the effective Hamiltonian:
\begin{equation} \label{renorm} \widetilde{\cH}=\cH_\text{sys} +
  \cV^\dagger \left( E - \cH_\text{env} \right)^{-1} \cV
\end{equation}
with the same eigenvalue $E$. Indeed, substituting $| \psi \ra = (\la
\psi_\text{sys} |, \la \psi_\text{env} | )^\dagger$ into $\cH | \psi
\ra = E | \psi \ra$, one gets $\cH_\text{sys} | \psi_\text{sys} \ra +
\cV^\dagger | \psi_\text{env} \ra = E | \psi_\text{sys} \ra $ and
$(\cH_\text{env} - E) | \psi_\text{env} \ra = - \cV | \psi_\text{sys}
\ra $, from which Eq.~(\ref{renorm}) follows.

Clearly, in the one-particle picture, the effect of the environment
can be presented as an (energy-dependent) renormalization of the bare
Hamiltonian of the system proper. One can use this systematic
procedure to estimate the strength of both the intra-chain and
environment's perturbation to the $pp\sigma$ bonding within the
chain. In doing so, below, we fix the energy $E$ at the gap center.

The contribution of $s$- or $d$-orbitals to the renormalization of the
$pp\sigma$ transfer integrals can be estimated by using a perturbative
expansion of the exact Eq. (\ref{renorm}):
\begin{equation} \label{renorm_approx}
  \widetilde{\cH}_{ij}=\cH_{\text{sys}, ij} +\sum_\alpha
  \frac{t_{i\alpha}t_{\alpha j}}{E-\epsilon_\alpha}
  -\sum_{\alpha\beta} \frac{t_{i\alpha}t_{\alpha\beta}t_{\beta
      j}}{(E-\epsilon_\alpha)(E-\epsilon_\beta)} +\ldots,
\end{equation}
where the Latin indices label intra-chain $p_z$ orbitals and the Greek
indexes label the rest of the orbitals. The small parameter
$t/(E-\epsilon)$ is indeed small, and the more so the further the
orbital energy $\epsilon$ from the gap center.

The essential elements of the Fock matrix are listed in Table
\ref{tab.FockM}.  Note that $d$-orbitals are also included in PM6
parametrization. Despite the large value of the pertinent transfer
integrals, the direct contribution of these orbitals to the
renormalization is at most a few percent because the orbitals are
almost empty. One infers from Table~\ref{tab.FockM} that the dominant
contribution to As--As bonding stems from $pp\sigma$-integrals.  The
$pp\pi$-integrals are four times smaller than $pp\sigma$, while the
$ss$-integrals are essentially negligible. Although $s$-orbitals do
not by themselves contribute significantly to the bonding in the
chain, they provide the main contribution to the renormalization of
the $pp\sigma$ transfer integrals, specifically by \emph{weakening}
the secondary bonding. This contribution, provided in the last column
in Table~\ref{tab.FockM}, turns out to be well approximated by the
expressions:
\begin{equation} \label{renorm_s} \Delta\epsilon\approx
  -t_{sp}^2/\epsilon_s,\quad \Delta t\approx
  2t_{sp}t_{sp}^\text{on-atom}/\epsilon_s,
\end{equation}
within the error of 20\% or less compared with the more accurate
Eq.~(\ref{renorm_approx}). Here
\begin{equation}
t_{sp}^\text{on-atom}\approx -g_{sp} P_{sp}/2,
\end{equation}
where $g_{sp}$ is the Coulomb integral for $s$ and $p$ orbitals of the
same atom (6 eV for As) and $P_{sp}$ is the corresponding entry of the
density matrix, which is the measure of actual hybridization. If
$P_{sp}>0$ - as is the case for AsH$_2$ chain - then the $sp$-mixing
weakens the secondary bonding, the magnitude of the effect proportional to
the hybridization strength.

%


\onecolumngrid

\newpage

\begin{center} \LARGE \textbf{Supplementary Material} \end{center}

\begin{figure}[h]
\centering
\includegraphics[width=\linewidth]{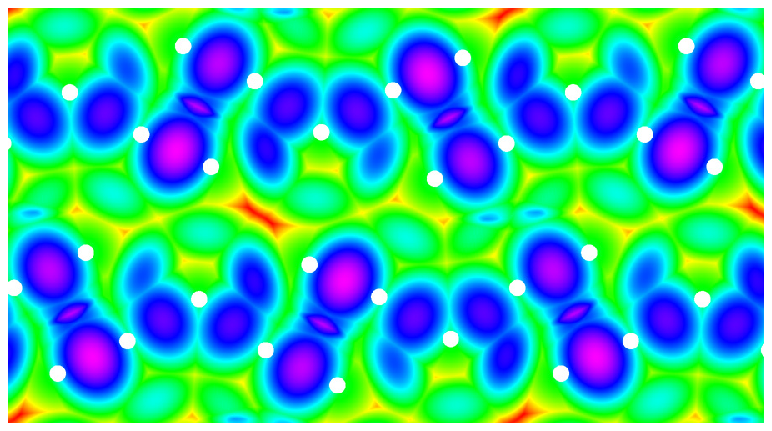}
\caption{Voids in As$_2$Se$_3$ crystal. The graph is generated as
  follows: First, we compute the distance $r_\text{min}$ to the
  nearest atom in each point $(\xi, \eta, \zeta)$ in the lattice
  coordinates. (The lattice is monoclinic.) Call the resulting
  function $r_\text{min}(\xi, \eta, \zeta)$.  Next, we make a
  projection of this function onto the plane of Fig.~9 of the main
  text, which is perpendicular to the $\zeta$ axis, according to the
  following recipe. For each point $(\xi, \eta)$, we vary $\zeta$ to
  find the largest value of $r_\text{min}( \xi, \eta, \zeta)$, call it
  $r(\xi, \eta)$, and assign a color to the point $(\xi, \eta)$ so
  that redder hues correspond to larger values of $r(\xi, \eta)$, more
  violet hues to smaller values of $r(\xi, \eta)$. The function
  $r(\xi, \eta)$ varies in the $[1.64, 2.52]$ \AA\ range, while the
  sum of the covalent radii is 2.4 \AA. As a result, intensely red
  regions mark voids in As$_2$Se$_3$ crystal. Atoms are marked by
  white balls.  The largest void is distorted octahedral; it is
  centered at point $(1/2,1/2,1/2)$ with Se atoms as vertices. }
\label{fig.As2Se3voids}
\end{figure}

\begin{figure}[h]
\centering
\fbox{\includegraphics[width=0.5\linewidth]{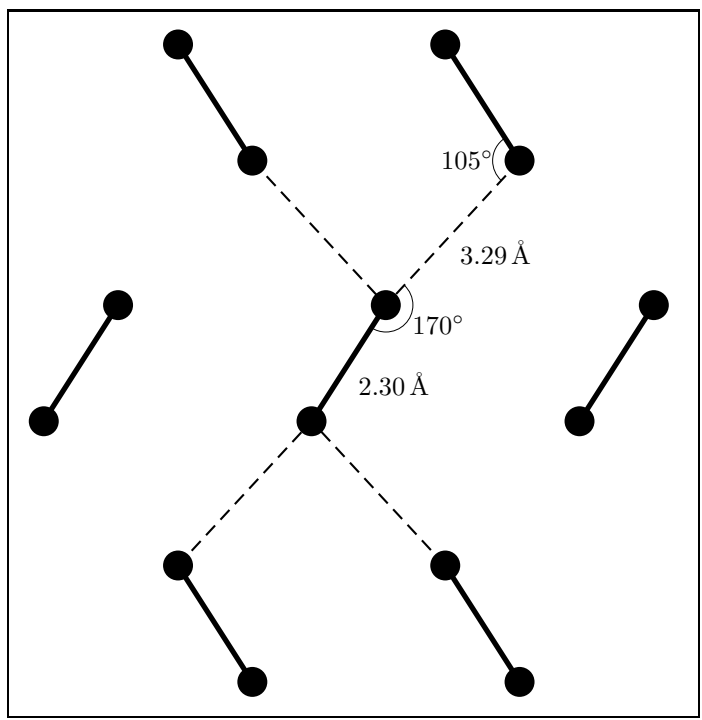}}
\caption{Fragment of a slice of bromine crystal. Compare the bond
  angles with those in Fig. 3 of the main text.}
\label{fig.Brcrystal}
\end{figure}

\begin{figure}[h]
\centering
\includegraphics[width=\linewidth]{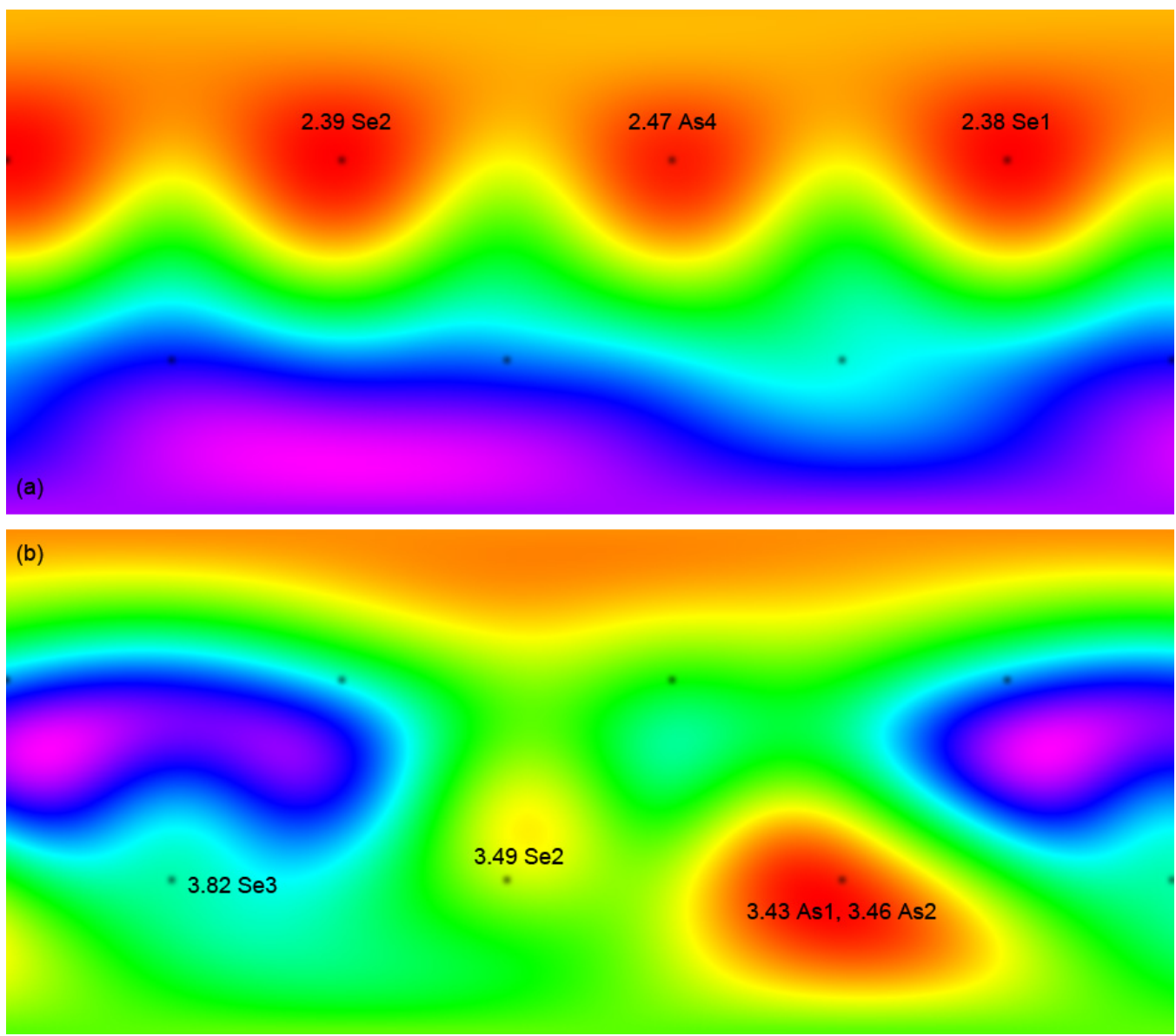}
\caption{Illustration of distorted octahedral coordination of the
  arsenic atom in the position As1 of As$_4$Se$_4$ crystal. Draw a
  sphere around the As1 atom at its covalent radius (1.2 \AA). For
  each point, compute the logarithm of the sum of the radial parts of
  the $p$ orbitals on the atoms within 4.5~\AA.  Assign a color to the
  value of this function so that redder hues correspond to larger
  values and more violet hues to smaller values. The resulting map is
  Merkator-projected on a rectangular area, where the top edge
  corresponds to the north pole. The resulting map is shown in
  panel (a). In perfect octahedral coordination, red areas would be
  centered at the black dots. (The line connecting the poles coincides
  with a C$_3$ axis of the octahedron.)  In panel (b), the nearest
  neighbors are excluded, to specifically highlight back-bonding.
  The pertinent neighbors are indicated, together with their distance
  to the As1 atom, in Angstroms.}
\label{fig.coo_As4Se4_As1}
\end{figure}

\begin{figure}[h]
\centering
\includegraphics[width=\linewidth]{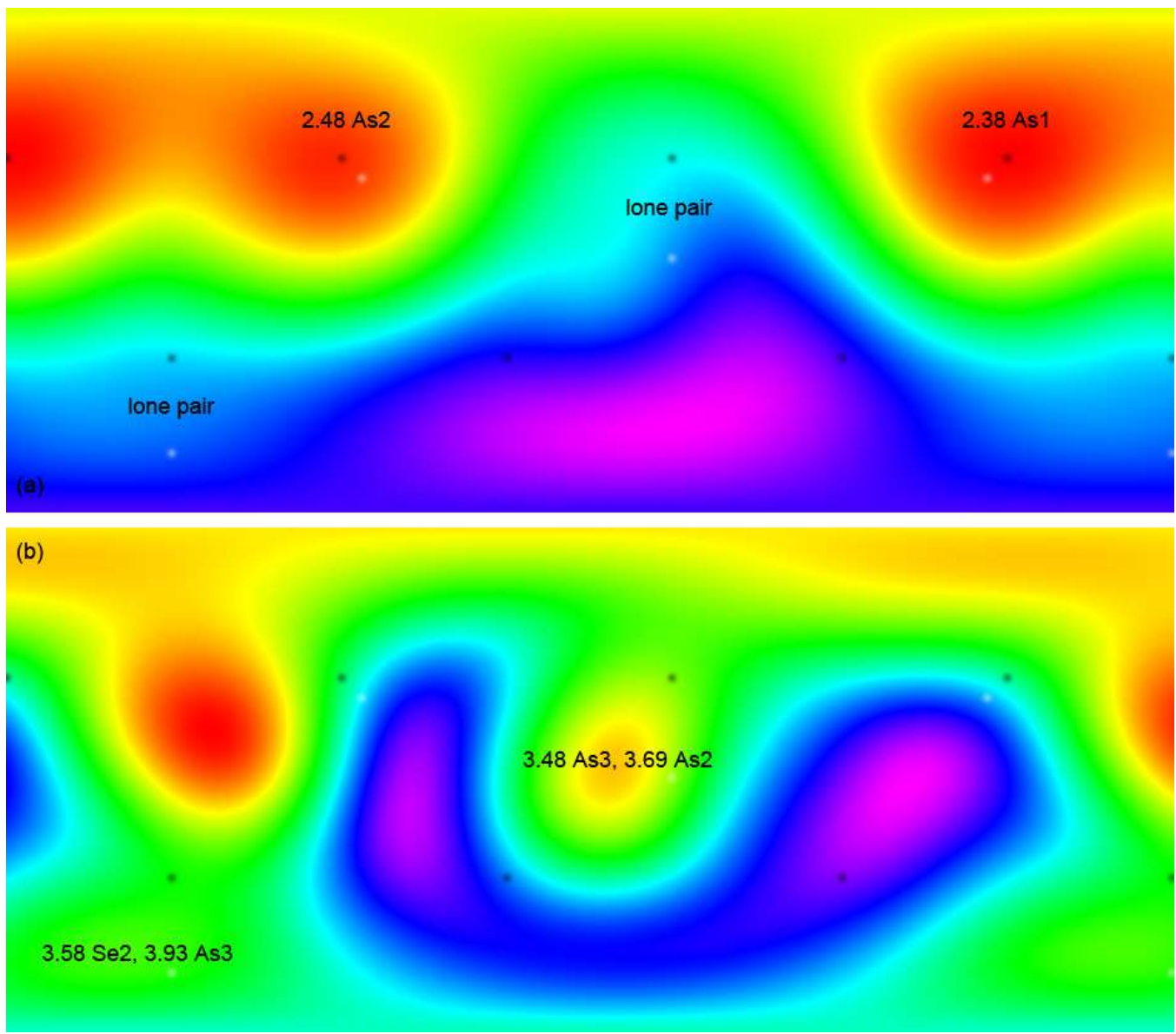}
\caption{Distorted tetrahedral coordination of the selenium atom in the position Se1 of As$_4$Se$_4$ crystal. See Fig.~\ref{fig.coo_As4Se4_As1} for explanation. Black and white dots indicate octahedral and tetrahedral coordination respectively.}
\label{fig.coo_As4Se4_Se1}
\end{figure}

\begin{figure}[h]
\centering
\includegraphics[width=\linewidth]{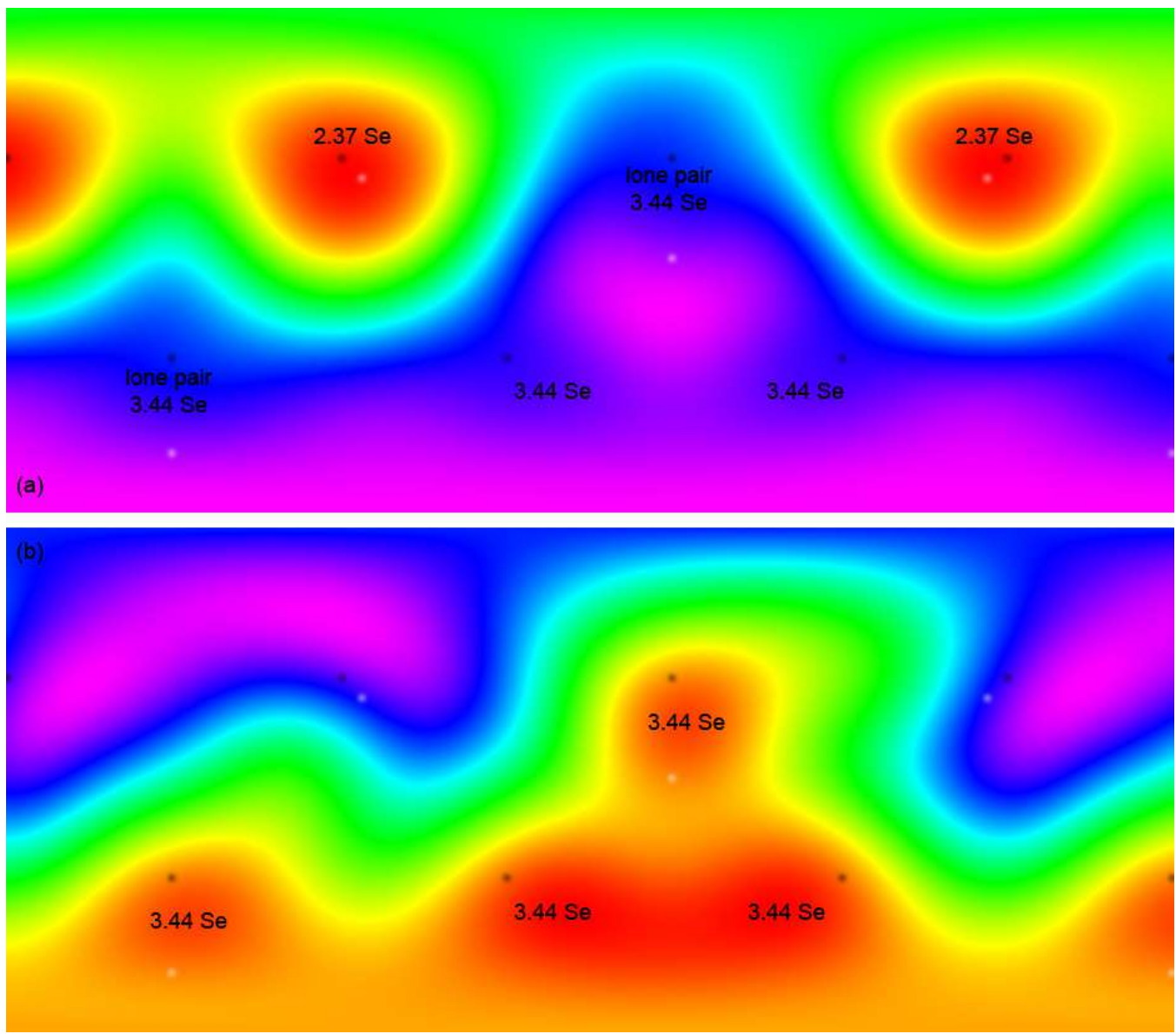}
\caption{Distorted octahedral coordination of the atoms in trigonal-Se crystal. See Figs.~\ref{fig.coo_As4Se4_As1} and ~\ref{fig.coo_As4Se4_Se1} for explanation.}
\label{fig.coo_tSe}
\end{figure}

\end{document}